\documentclass[review]{elsarticle}
\makeatletter
\def\ps@pprintTitle{%
 \let\@oddhead\@empty
 \let\@evenhead\@empty
 \def\@oddfoot{\centerline{\thepage}}%
 \let\@evenfoot\@oddfoot}
\makeatother
\usepackage{lineno,hyperref}
\modulolinenumbers[5]
\usepackage{gensymb}

\usepackage{graphicx}
\usepackage{subfigure}
\usepackage{booktabs}
\usepackage{longtable}
\usepackage{pdflscape} 
\usepackage{amsmath}
\newtheorem{definition}{Definition}
\newtheorem{properties}{Properties}
\newtheorem{proposition}{Proposition}

\usepackage{natbib}

\begin{document}

\begin{frontmatter}

\title{A Spatio-Temporal Model for Predicting Wind Speeds in Southern California}

\author[first,second]{Mihaela Puica\corref{cor}}
\ead{mihaela.puica@lseg.com, mihaelp@student.matnat.uio.no}
\address[first]{Refinitiv Commodities Content \& Research}
\address[second]{Department of Mathematics, University of Oslo}


\author[second]{Fred Espen Benth}


\cortext[cor]{Corresponding address: Refinitiv (part of London Stock Exchange Group) Dronning Eufemias gate 16, N-0191, Oslo, Norway}


\begin{abstract}
The share of wind power in fuel mixes worldwide has increased considerably. The main ingredient when deriving wind power predictions are wind speed data; the closer to the wind farms, the better they forecast the power supply. The current paper proposes a hybrid model for predicting wind speeds at convenient locations. It is then applied to Southern California power price area. We build random fields with time series of gridded historical forecasts and actual wind speed observations. We estimate with ordinary kriging the spatial variability of the temporal parameters and derive predictions. The advantages of this work are twofold: (1) an accurate daily wind speed forecast at any location in the area and (2) a general method applicable to other markets.
\end{abstract}


\end{frontmatter}

\linenumbers

\section{Introduction}

Wind generation has become an increasingly important production technology in nearly all power markets. Statistics from the International Renewable Energy Agency (IRENA) show that the amount of wind power installed capacity worldwide has more than doubled since 2010, reaching 622.4 GW at the end of 2019 \footnote{IRENA reports an installed capacity of 180.8 GW in 2010 and 622.4 GW in 2019. See: \url{https://www.irena.org/wind}}. This technology has potential to cover 35\% of world's power needs and in order to meet climate targets, the installed capacity must reach 6000 GW by 2050 ~\cite{IRENAReport}. From this perspective, accurate wind power predictions are paramount for day-ahead and real-time power clearing (\cite{LangeFocken}, \cite{Hodgeetal}) or for regulation volumes on the balancing power market (\cite{Leietal}, \cite{Botterudetal}). In exchange, these help increasing the wind power penetration itself as the power producers can plan their generation better and avoid imbalance costs.

The present paper examines in an original way the core component of a wind generation forecast, namely the wind speed forecast. The aim of this research is to build a method of predicting day-ahead wind speeds in a wide range of topographies and in locations where high installed capacities are reported. This answers an important market need and is in line with the research directions suggested by~\cite{Costaetal} and~\cite{Chang}. Starting from a random field under grid format, we derive parameters of a stochastic model for time predictions in the spirit of \cite{BenthBook} and \cite{SaltyteBenthBenth}. Afterwards, we apply ordinary kriging to determine spatially-relevant parameters in places outside the grid points. With the new parameters thus obtained, we apply the same stochastic model to derive forecasts in time at any new site. We use airport locations in order to validate the model as we have actual wind speed observations from the Meteorological Aerodrome Reports (METAR). The novelty of this approach is given by three aspects.

Firstly, the paper addresses the issue of wind speed downscaling in complex terrains in an original fashion. We work with time series of gridded measurements that we eventually shift in space via an ordinary kriging estimate of the temporal parameters. Thus we are moving from a coarser resolution (i.e. 0.5\degree$\times$0.5\degree) to a denser one, a process called downscaling (see, for example,~\cite{Giebeletal}). This can be of two main classes: dynamical or statistical. The former is based on regional climate models of the denser area (\cite{Lebassi-Habtezionetal},~\cite{Loetal}). The latter implies building statistical relationships between the two data sets, be it in the form of regression (\cite{Gutierrezetal},\cite{Pryoretal}), Kalman filters (\cite{CassolaBurlando}), neural networks (\cite{Sailoretal}) or statistical transforms from one dataset to the other (\cite{Michelangelietal}, \cite{TangBassill}). The application of ordinary kriging to the study of spatial variability of temporal parameters belongs to the second category. To the best of our knowledge, this procedure is new in the field of meteorological downscaling. We note that \cite{HaslettRaftery} employed ordinary kriging in a similar problem. However, this estimate has been performed on the resulting residuals of the temporal model. Also, the temporal behaviour has been exploited \textit{a posteriori} while we propose the opposite. 

Secondly, the use we make of the available data on the field is rather unique. While each wind farm operator may have \textit{in situ} wind speed observations from local anemometers, the dissemination of such data to market or academic actors is limited. Therefore, modelling entire market areas becomes challenging. With this work, we propose innovations on two levels. On the one hand, we build a spatio-temporal random field based on gridded wind speed predictions from the operational forecasts of the European Centre for Medium-Range Weather Forecast (ECMWF). These will henceforth be denominated EC forecasts. They update every 6 hours and we take the first 6 hours of each forecast horizon to build time series in each of the 0.5\degree$\times$0.5\degree grid points. We treat these data as 'actual measurements' and we use them to train our models. Indeed, it is recognized that by doing so we already account for any potential systematic errors in the weather predictions (\cite{Akylasetal}, \cite{Beyeretal}). On the other hand, we validate our spatio-temporal model based on weather measurements from METAR. The locations where these reports are available are usually airports. We have chosen conveniently the closest airports to wind farms. Our case study is designed for Southern California price zone area. However, the model hereby developed is applicable to other market areas as well.

Thirdly, our model is a hybrid of three of the four recognized techniques for wind speed forecasting (\cite{Costaetal}, \cite{Leietal}, \cite{Chang}): physical, conventional statistical, spatial statistical and AI. The first three methods are encompassed to some extent in our model. The physical methods are based on atmospheric considerations and use numerical weather predictions (eg: \cite{PalomaresCastro}). The operational EC forecasts from ECMWF used for training our model belong to this category. The conventional statistical models refer, for example, to AR/ARMA/ARIMA models or Kalman filters. In the current method, we find AR(2) to model well the temporal behaviour in the grid points, in line with the findings of \cite{Nfaoui}.  We model autoregressively the wind speed resultant instead of the two $U,V$ components (i.e. $\sqrt{U^2+V^2}$). This is emphasised in \cite{ErdemShi} as a better technique for wind speed forecasting. Our approach also encompasses spatial statistical techniques, by way of ordinary kriging. To the best of our knowledge, this technique has not been applied on model parameter variability before. Instead, other spatial statistical techniques have been employed. For instance, \cite{Damousisetal} uses a constant delay method based on the idea that wind speeds propagate from one site to another by a function of $\Delta \tau$. Other factors have been used as inputs of the spatial variability, as for example, the distance between sites (\cite{Corotisetal}), the topographical elevation difference (\cite{Beyeretal1993}) or the difference between the two-dimensional U-V components (\cite{PalominoMartin}).

The structure of the paper revolves around the three aforementioned aspects. In Section~2 we discuss the problem formulation in more detail, including a visual account of what we want to achieve, specificities of the data and the terrain, as well as their descriptive statistics. Section~3 introduces the model and adds a technical part on the space-time random field. We discuss in general the AR model first and afterwards the semivariogram and ordinary kriging. In Section~4 we analyse the results and we determine benchmarks both in-sample and out-of-sample. We eventually present our concluding remarks in Section~5.

\section{Space-time problem}\label{space-time problem}

\begin{figure}
\centering
  \includegraphics[width=0.9\textwidth]{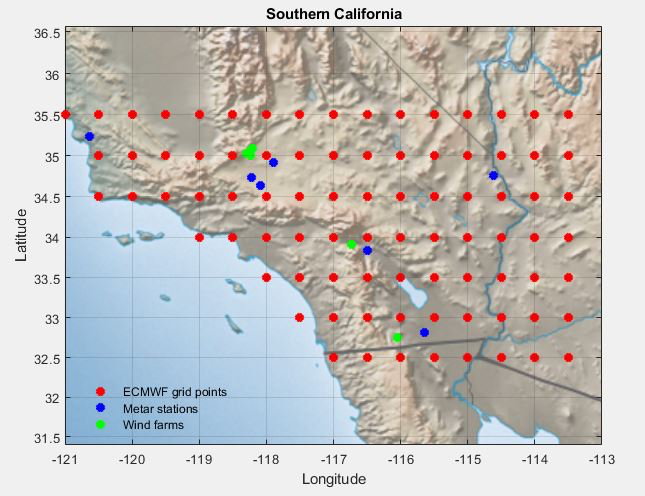}
  \caption{Map of Southern California. The green points are the biggest wind farms in the area (Alta Wind Eenergy Center, San Gorgonio Path Wind Farm, Ocotillo wind energy project). The red points are the ECMWF forecasted wind speeds. The blue points are the airports covered by METAR data (KEDW, KWJF, KPMD, KSDB, KEED, KPSP, KNJK).}
  \label{map}
\end{figure}

With our model, we aim at predicting day-ahead wind speeds at the wind farm locations (marked in green in Fig~\ref{map}). Since there are no available historical measurements at those sites, we will train our model on data nearby. 

In the grid points (marked in red), we have past operational forecast data from ECMWF\footnote{Data originating from ECMWF and processed by Refinitiv}. These forecasts are released every 6 hours and span over a grid of 0.5\degree$\times$0.5\degree \ resolution\footnote{As a rule of thumb, 1\degree $\approx$ 111 km. This varies due to the geoidal shape of the Earth. In our study, we use the haversine formula to determine the great-circle distance between any 2 points.}. From each forecast, we use the shortest horizon until the next one is released. For example, from the 00 (UTC) forecast, we use the values predicted for 00.00-05.00, from the 06 (UTC) forecast we use the 06.00-11.00 values and so on. By doing this, we end up with a time series of forecasted wind speeds that we regard as 'actual measurements'. We therefore assume their high accuracy on the short-term. Furthermore, the forecasts are made for 100m height above ground level (AGL). This is more relevant to the wind power production as it is known that optimal windmill hub heights are in the range of 75-80 m for onshore farms \cite{Leeetal}. Moreover, each forecast covers a $U$- and $V$- parameter. They are the eastward and the northward horizontal component, respectively, and they are measured in m/s. In the sequel, we will work with the wind speed resultant (i.e. $\sqrt{U^2+V^2}$), as suggested by \cite{ErdemShi}.

The METAR datasets\footnote{Data publicly available at \url{https://www.aviationweather.gov/metar} collected by Refinitiv} provide us with the wind speed actual observations around airports (marked in blue). We selected conveniently the closest airports to the wind farms. The data are reported in knots\footnote{1 kt $\approx$ 0.514444 m/s} and measure wind speeds at 10m AGL. The METAR datasets will be used for model validation in space and time. However, given that the model will be trained on observations at 100m AGL, we will use an approximation log wind profile law in order to scale up our METAR data.  In practice, we will use a simplified version:

\begin{equation}\label{loglaw}
ws_2(z_2) = ws_1(z_1) \frac{\log \big(\frac{z_2}{z_0}\big)}{\log \big(\frac{z_1}{z_0}\big)}
\end{equation}
where $ws_2, ws_1$ are the wind speeds at heights $z_2$ and $z_1$. $z_0$ is a roughness length coefficient that depends on the terrain. Given the nature of the METAR observations, the terrain will be mainly airport runways. Therefore, we will assume that $z_0=0.0024$\footnote{According to the Danish Wind Industry Association. 2003 Wind Energy Reference Manual,  accessed in January 2020, available at: \url{http://drømstørre.dk/wp-content/wind/miller/windpower\%20web/en/stat/units.htm}}.

\subsection{The Data}

\begin{figure}[htbp]
  \includegraphics[scale=0.44]{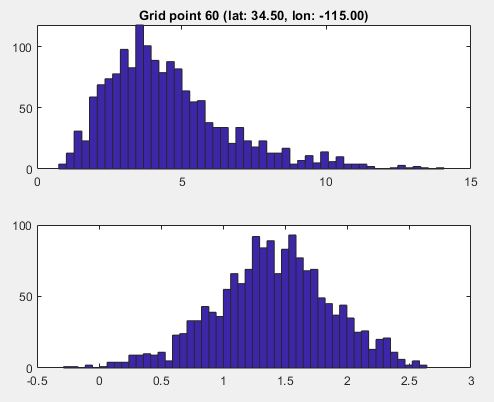}
  \includegraphics[ scale=0.44]{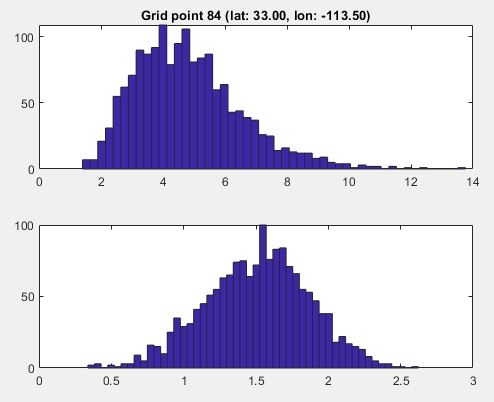}
 \caption{Histograms of daily wind speeds in 2 grid points: raw data (upper chart) and logarithmised data (lower chart)}
  \label{logarithmizedwindspeeds}
\end{figure}

The data derived as explained earlier are averaged up to daily resolution. We work with a sample of wind speeds from 1 February 2015 until 1 July 2019. We extend this between 1 July 2019 and 1 Mar 2020 for model validation purposes. All in all, we have 1612 in-sample data points. 
All the daily averaged wind speed values are transformed via a simple Box Cox transformation, namely by taking logarithm. This is deemed important for symmetrizing data, in particular for wind speeds (\cite{BenthBook}). Indeed, Figure~\ref{logarithmizedwindspeeds} illustrates this principle.

The time series of forecasted wind speeds cover 85 grid points, spanning from the coastal areas to the border with Mexico (south), with Arizona (south-east) or Nevada (east).
The actual observations from the METAR reports cover 21 airports across Southern California. We select 7 observation sites, 6 of which are in the proximity of important wind farms (see from Fig~\ref{map}):
\begin{itemize}
\item Edwards Air Force Base (KEDW), Lancaster Fox Airfield (KWJF), Palmdale (KPMD), Sandberg (KSDB) close by Alta Wind Energy Center in the Mojave Desert
\item Palm Springs Regional Airport (KPSP) close by San Gorgonio Path Wind Farm
\item El Central Naval Air Facility (KNJK) near Ocotillo wind project
\item Needles Airport (KEED) near Colorado River
\end{itemize}

\begin{figure}[htbp]
 
 \centering
    \subfigure[]{\includegraphics[width=0.45\textwidth]{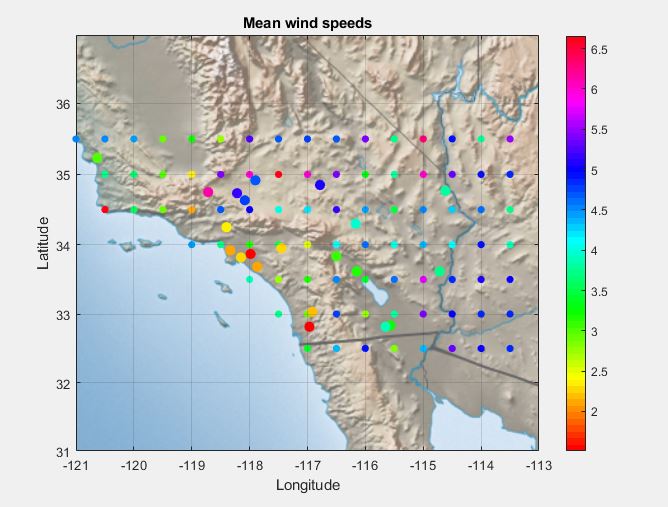}} 
      \subfigure[]{\includegraphics[width=0.45\textwidth]{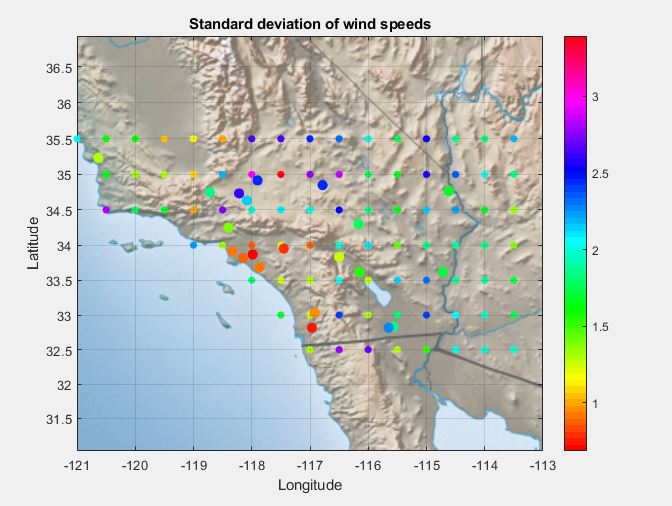}} 
    \subfigure[]{\includegraphics[width=0.45\textwidth]{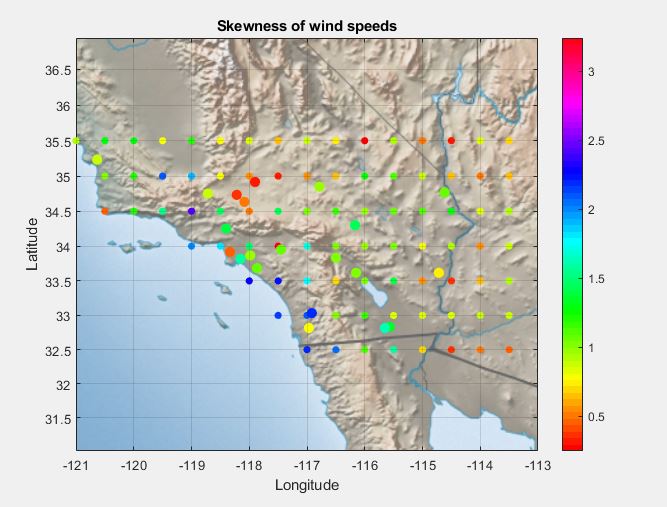}} 
      \subfigure[]{\includegraphics[width=0.45\textwidth]{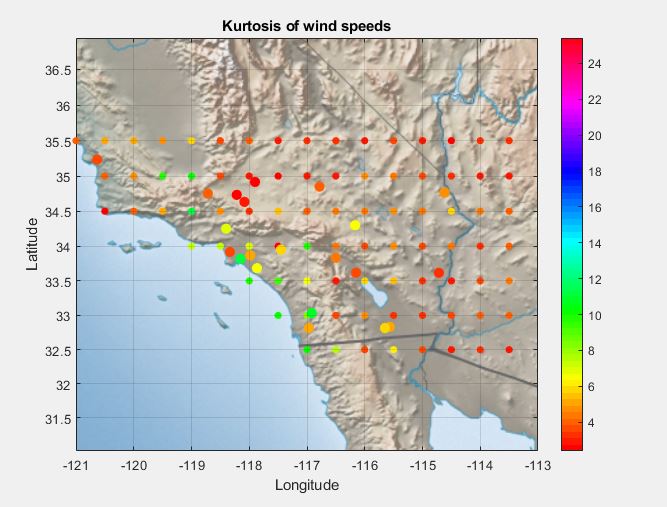}} 

 \caption{Descriptive statistics of the gridded (small points) and Metar time series (big points)}
  \label{descriptivestats}
\end{figure}

Fig~\ref{descriptivestats} depicts the main descriptive statistics for the time series in each site. All the inputs are detailed in Table~\ref{gridpointsstatistics}. We remark three key aspects.

\begin{figure}[htbp]
 \centering
    \subfigure[]{\includegraphics[width=0.440\textwidth]{Figure3a.jpg}} 
    \subfigure[]{\includegraphics[width=0.45\textwidth]{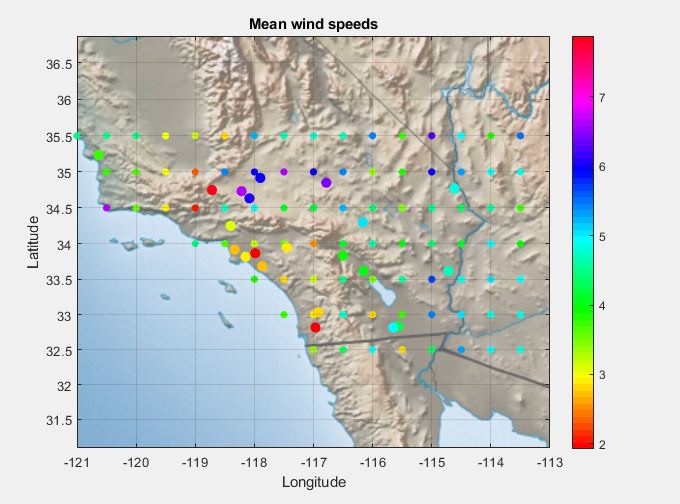}} 
 
 \caption{Average levels before (a) and after the log-wind profile law (b)}
  \label{scalingmetar}
\end{figure}

Firstly, there is an important spatial variability of the averaged wind speeds in the Mojave Desert area, located approximately around 35\degree \ latitude and -118\degree \ longitude. However, the same does not entirely hold for the actual observations in the same area. This shows that numerical weather predictions in such a complex terrain can be rather challenging. 
The second aspect is related to the positive skewness and kurtosis for both the grid and the airport sites. This suggests that the logarithmic transformation illustrated in Fig~\ref{logarithmizedwindspeeds} is applicable to all the locations as it reduces the right-skewness.
The third key point refers to the differences in mean level between grid and airport sites located nearby. This is caused by the fact that the datasets from the two sources express wind speeds at different heights, hence the need for a transformation. We recall that the gridded data represent wind speeds at 100m AGL while the METAR reports reflect this at 10m AGL. Upon applying the log wind profile law from Eq~\ref{loglaw}, the average levels become comparable (see Fig~\ref{scalingmetar}).

\section{The Model}

Let $D$ be the spatial domain of the grid points. Any $s_k \in D$ with $k=\{1, 2 \dots 85 \}$ contains a set of coordinates (latitude, longitude). Also, let $D^\prime$ be the domain of the relevant airports in Southern California, with any $s_j \in D^\prime$, $j=\{1,2,\dots 6\}$.
In addition, define $T$ as the time domain, in our case, of all the days between 1 February 2015 and 1 July 2019. 

We define a random field $\{W(s_k,t)=\log(Z(s_k,t)) | s_k\in D, t\in T\}$ describing the logarithm of all the wind speeds in the grid points from Fig~\ref{map}. 

We will introduce first a temporal model for $W(s_k,t)$, with its associated $\Theta_k$, the set of all the model parameters. This will be done in stages, following the method from \cite{BenthBook} \cite{SaltyteBenthBenth}, \cite{SaltyteBenthSaltyte}.

Afterwards, we will describe the ordinary kriging method of transitioning from $\Theta_k$ to $\Theta_j$ following \cite{CressieWikle}, where $\Theta_{\cdot} $ is the set of parameters corresponding to the temporal model.

Now, on the temporal domain, we will model the wind speeds as a stochastic process of the form:
\begin{equation}\label{totalmodel1}
W(s_k,t)=S(s_k,t)+A(s_k,t)+\varepsilon(s_k,t), \forall s_k\in D
\end{equation}
where $S(s_k,t)+A(s_k,t)$ is the mean component and it is a deterministic function of time while $\varepsilon(s_k,t)$ is a residual process random in time. In particular, we choose:

\begin{equation}\label{seasonalityeq}
 S(s_k,t)=a_0^k+\sum_{i=1}^6  \bigg [ a_{2i-1}^k \cos \bigg(\frac{2\pi i t}{365.25} \bigg)+a_{2i}^k \sin\bigg(\frac{2\pi i t}{365.25} \bigg) \bigg ]
\end{equation}
the stationary mean of $W(s_k,t)$, a Fourier sine and cosine series truncated to 6 terms, where we assume no linear trend;

\begin{equation}\label{ar(2)}
A(s_k,t)=\alpha_1^k\big [W(s_k,t-1)-S(s_k,t-1)\big]+\alpha_2^k \big [W(s_k,t-2)-S(s_k,t-2)\big ]
\end{equation}
an AR(2) process;

\begin{equation}\label{epsilon}
\varepsilon(s_k,t)=\sqrt{b_0^k+b_1^k  \cos \bigg(\frac{2\pi t}{365.25} \bigg)+b_2^k  \sin\bigg(\frac{2\pi t}{365.25} \bigg)} \cdot \epsilon(s_k,t)
\end{equation}
 a seasonal volatility function and $\epsilon \sim \mathcal{N}(0,1)$ a temporally independent random variable. 

 In this model, the set of fitted parameters is 
$$\Theta_k = \{ a_0^k, a_1^k, \dots a_{12}^k, \alpha_1^k, \alpha_2^k, b_0^k, b_1^k, b_2^k\}, \forall k\in \{1, 2, \dots 85\}$$
and it will be determined upon estimation of the three processes. Note that the superscript $k$ stands for the $k$-th grid point.

From a spatial point of view, we can see $\{W(\cdot,t) \}$ as a partially sampled stochastic process, satisfying a decomposition as in Eq~\ref{totalmodel1}. This has a set of 85 observations, each with an associated $\Theta_{\cdot}$, as described earlier. Let us change the spatial domain and take any $s_0 \in D^\prime$. Our goal is to determine the hidden process $W(s_0,t)$ which we also choose to decompose as in Eq~\ref{totalmodel1}. More precisely, we want to determine a modelling-error free version of $W(s_0,t)$ since we know that our decomposition of the observations $W(s_k,t)$ carries its intrinsic errors. We write this as
$$W^\prime(s_0,t)=S^\prime(s_0,t)+A^\prime(s_0,t)+\varepsilon^\prime(s_0,t)$$

The ordinary kriging predictor for each separate term is of the form (\cite{Cressie}) :
\begin{equation}\label{krigingS}
\hat{S^\prime}(s_0,t)= \sum_{k=1}^{85} \lambda_k S(s_k,t)
\end{equation}
\begin{equation}\label{krigingA}
\hat{A^\prime}(s_0,t)= \sum_{k=1}^{85} \nu_k A(s_k,t)
\end{equation}
\begin{equation}\label{krigingepsilon}
\hat{\varepsilon^\prime}(s_0,t)= \sum_{k=1}^{85} \mu_k \varepsilon(s_k,t)
\end{equation}

Replacing the actual forms of $S(s_k,t), A(s_k,t), \varepsilon(s_k,t)$ from Eq~\ref{seasonalityeq}-\ref{epsilon} and admitting that there is a "true" similar decomposition for $S^\prime(s_0,t), A^\prime(s_0,t), \varepsilon^\prime(s_0,t)$, we translate the ordinary kriging equations for each spatial term:

\begin{align}\label{kriging}
\hat{a}_i(s_0)&:=\hat{a_i^0} =\sum_{k=1}^{85}\lambda_k a_i^k \ \ \ \text{such that} \sum_{k=1}^{85}\lambda_k=1, \ \ \ \ i=0,1,2 \dots 12 \nonumber \\
\hat{\alpha}_i(s_0)&:=\hat{\alpha_i^0} =\sum_{k=1}^{85}\nu_k \alpha_i^k, \ \ \text{such that} \sum_{k=1}^{85}\nu_k=1, \ \ \ \  i=1,2  \\
\hat{b}_i(s_0)&:=\hat{b_i^0} =\sum_{k=1}^{85}\mu_k b_k, \ \ \text{such that} \sum_{k=1}^{85}\mu_k=1, \ \ \ \ i=0,1,2 \nonumber
\end{align}

Therefore, our problem reduces to finding the optimal predictors of $\lambda, \nu, \mu$ for each parameters in $\Theta_0$, meaning those that (1) yield the minimum mean-squared error and (2) are unbiased.
The solutions to the first optimization in Eq~\ref{kriging} is given by the ordinary kriging theory (\cite{Cressie},\cite{CressieWikle}) as the solution to the system of 86 equations:

\begin{equation}\label{systemkriging}
\Gamma \boldsymbol{\lambda}=\boldsymbol{\gamma}
\end{equation}
where
\begin{align}\label{solutionkriging}
\boldsymbol{\lambda} &= \bigg( \lambda_1, \lambda_2, \dots \lambda_{85}, m\bigg)\prime \nonumber \\ 
\boldsymbol{\gamma} &= \bigg(\gamma(s_1-s_0),\gamma(s_2-s_0), \dots \gamma(s_{85}-s_0), 1\bigg)\prime \\
\Gamma &= \left(\begin{array}{ccccc}  \nonumber
   \eta&\gamma(s_2-s_1)&\dots & \gamma(s_{85}-s_1) & 1\\ 
    \gamma(s_2-s_1)& \eta &\dots & \gamma(s_{85}-s_2) & 1 \\
    \vdots&\vdots &\ddots& \vdots& \vdots \\
   \gamma(s_{85}-s_1) & \gamma(s_{85}-s_2) & \dots &\eta & 1\\
   1 & 1& \dots & 1 & 0 \end{array}\right) 
\end{align}

Here, $m$ is a Lagrangian multiplier used in the optimization process to ensure unbiasedness (i.e. $\sum \lambda_j=1$). 

$\gamma(\cdot)$ is the semivariogram of the spatial distribution of $a_i$ for $ i=0,1, \dots 12$. In addition, $\eta=\gamma(0)$ is the nugget effect of the variogram or the modelling error of fitting $a_i$ to each of the 85 time series of gridded data. In other words, it is the variance of the fitting error.

 For the kriging optimal solution to exist, we assume that the processes $a_i$, $i=0,1,\dots 12$ are intrinsically-stationary (\cite{Cressie}).

\subsection{Semivariogram}

In the sequel, $Y$ stands for all the processes fitted earlier for their temporal behavior: $a_0,a_1,a_2,\dots a_{12},\alpha_1, \alpha_2, b_0,b_1,b_2 $.

The semivariogram was first introduced by Matheron in 1963 \cite{Matheron}. \begin{definition}
The theoretical semivariogram  of a process $Y$ is the function $\gamma: D\to \mathbf{R}$ defined as
$$\gamma(s_n-s_m)=\frac{1}{2}Var(Y(s_n)-Y(s_m)), \forall s_n, s_m \in D $$
\end{definition}

\begin{properties}\label{props} The following hold for the theoretical semivariogram:
\begin{enumerate}[(i)]
\item $\gamma(-h)=\gamma(h)$
\item $\gamma(0)=0$
\item If  $\lim_{h\to 0}\gamma(h)=C_0>0$, then $C_0$ is called the nugget effect
\item $\gamma$ must be conditionally negative semidefinite:
$$\sum_{i=1}^{85} \sum_{j=1}^{85} w_i w_j \gamma(s_i-s_j)\leq 0, \ \ \  \forall \{w_k\}_{k=1}^{85} \ \ \text{satisfying} \sum_{j=1}^{85}w_j=0$$
\item Let $C(h)=Cov(Y(s),Y(s+h))$ be the covariance function, then it holds
$$\gamma(h)=C(0)-C(h)$$
This relation makes the transition between the semivariogram and the covariance function. Note that $C(0)$ is called the sill of the semivariogram
\end{enumerate}
\end{properties}

Empirically, the semivariogram can be estimated from the data with the following formula:
\begin{equation}\label{semivariogramempirical}
\hat{\gamma}(h)=\frac{1}{|N(h)|}\sum_{N(h)}\bigg[ Y(s_i)-Y(s_j)\bigg]^2
\end{equation}
where $N(h)=\{(s_i,s_j)| \ \ ||s_i-s_j||=h\}$

In our case study we determine the empirical semivariograms and covariances of all the parameters $a_1,a_2,\dots a_{12}$, $\alpha_1,\alpha_2$, $b_0,b_1,b_2$. We seek theoretical semivariograms to fit our empirical results while making sure that they fulfill all the characteristics described by Properties~\ref{props}. Using some of the classical suggested functions (eg: \cite{CressieWikle}, \cite{CressieHuang}), we combine spherical, exponential, sine hole-effect or nested models. Results are depicted in Fig~\ref{fittedvariograms}.

\subsection{Ordinary Kriging}\label{OK}
Kriging can be performed under the assumption that the process $Y$ is intrinsically stationary \cite{Cressie}.
\begin{definition}A process $Y$ is intrinsically stationary if \begin{enumerate}[(i)]
\item $Y$ has a constant mean: $E\big [Y(s)\big ]=\mu, \forall s\in D$
\item The variance of increments is finite and only depends on the increments: $Var(Y(s_n)-Y(s_m))=2\gamma(s_n-s_m)<\infty$ and the semivariogram only depends on the distance between observations.
\end{enumerate}
\end{definition}

If we want to estimate the value of the process $Y$ whose mean we do not know in a site $s_0\in D\prime$ outside our grid, ordinary kriging gives that:
\begin{equation}\label{krigingestimator}
\hat{Y}(s_0)=\sum_{i=1}^{85}\lambda_i Y(s_i)
\end{equation}
where $\boldsymbol{\lambda}=\big( \lambda_1, \lambda_2, \dots \lambda_{85}\big)\prime$ is the solution to the following optimization problem

\begin{align}\label{optimizationproblem}
\min_{\boldsymbol{\lambda}} \ & E \bigg[ (Y(s_0)-\hat{Y}(s_0))^2\bigg] \nonumber \\
&\text{such that} \\
& \sum_{i=1}^{85}\lambda_i =1  \nonumber
\end{align}

The objective function can be written in short as (see \ref{derivation}):
\begin{equation}
E \bigg[ (Y(s_0)-\hat{Y}(s_0))^2\bigg]  = - \sum_{i=1}^{85}\sum_{j=1}^{85}\lambda_i \lambda_j \gamma(s_i-s_j)+ 2\sum_{i=1}^{85}\gamma(s_i-s_0)
\end{equation}

We introduce in the constraint a conveniently chosen Lagrange multiplier $L:=-2m$. Thus, we aim to minimize the function:

\begin{equation}\label{optmizationfunc}
f(\boldsymbol{\lambda})= - \sum_{i=1}^{85}\sum_{j=1}^{85}\lambda_i \lambda_j \gamma(s_i-s_j)+ 2\sum_{i=1}^{85}\gamma(s_i-s_0) + 2m \bigg( \sum_{i=1}^{85}\lambda_i-1\bigg)
\end{equation}
Taking partial derivatives of $f$ w.r.t. $\lambda_1, \lambda_2, \dots \lambda_{85}, m$ and equating to 0, we obtain the system of equations:

\begin{align}
&\sum_{j=1}^{85}\lambda_j \gamma(s_j-s_i)+m=\gamma(s_i-s_0), \ \ \ \ i=1,2,\dots 85 \\
&\sum_{j=1}^{85}\lambda_j=1 \nonumber
\end{align}

which eventually can be written as $\Gamma \boldsymbol{\lambda}=\boldsymbol{\gamma}$, with $\Gamma, \boldsymbol{\lambda}$ and $\boldsymbol{\gamma}$ given in  Eq~\ref{solutionkriging}.

Furthermore, the predictor $\hat{Y}(s_0)$ is unbiased by definition. Indeed, $$E\bigg [ Y(s_0)-\hat{Y}(s_0)\bigg]=E\bigg [ Y(s_0)-\mu -(\hat{Y}(s_0)-\mu)\bigg]=-\sum_{i=1}^{85}\lambda_i E[Y(s_i)]-\mu=0$$

Moreover, the mean-squared error of the optimal solution is:
$$\sigma^2(s_0)=Var(Y(s_0)-\hat{Y}(s_0))=E \bigg[ (Y(s_0)-\hat{Y}(s_0))^2\bigg]=\sum_{i=1}^{85}\lambda_i \gamma(s_i-s_0)+m-\eta$$
where $\eta$ is the nugget effect (i.e. $\eta=\gamma(0)$).

\section{Results}

Next, we will see how the model performs on our specific dataset from Southern California. We apply the model described earlier to predict wind speeds in 7 airport locations. This allows us to benchmark the model against METAR observations. Thus we assess the strengths and weaknesses of the model given the different terrains and geologies that we will be approaching.

First, we apply the temporal model described earlier on each of the 85 time series in the grid points. In the next step, we optimize the temporal parameters for any given location via ordinary kriging.

\subsection{Temporal model}

The temporal model described in Eq~\ref{totalmodel1} is fitted to the 85 time series with in-sample data from 1 February 2015 until 1 July 2019. 

\begin{figure}[htbp]
 \centering
    \subfigure[]{\includegraphics[width=0.45\textwidth]{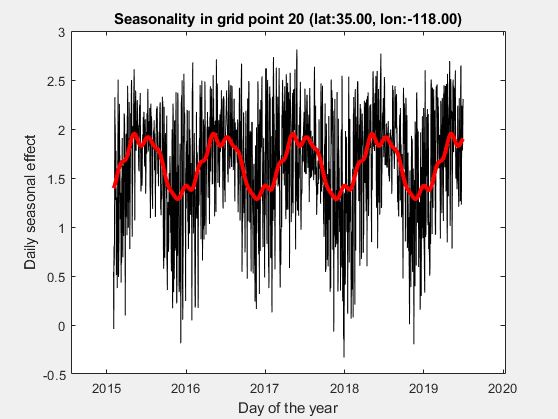}} 
      \subfigure[]{\includegraphics[width=0.45\textwidth]{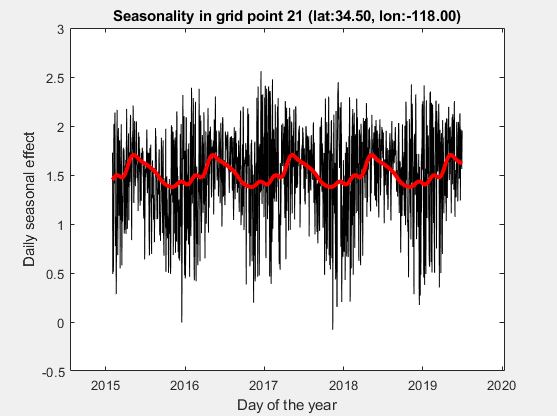}} 
    \subfigure[]{\includegraphics[width=0.45\textwidth]{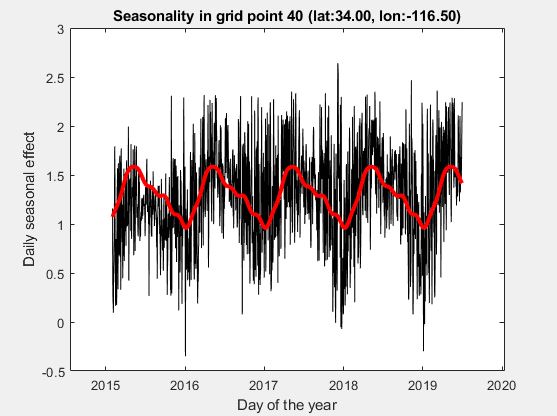}} 
      \subfigure[]{\includegraphics[width=0.45\textwidth]{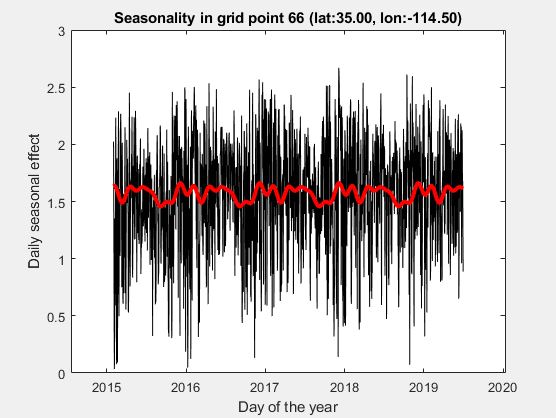}} 
  
 \caption{Seasonality effects for wind speeds in the Mojave Desert (a,b), Coachella Valley (c), Mojave Valley (d)}
  \label{seasonality}
\end{figure}

Initially, we study the seasonality effects on wind speeds and fit a 6-terms Fourier series. These correspond to the $S(s_k,t), k=1,\dots 85$ parameters. Results show that seasonality is clearer in the Mojave Desert area or in the Coachella Valley (Fig~\ref{seasonality} (a), (b), (c)) and more erratic towards the head of the Mojave Valley (Fig~\ref{seasonality}, (d)). Also, the summer-winter cycle seems to be more pronounced across the coastline rather than inland.

\begin{figure}[htbp]
 \centering
    \subfigure[]{\includegraphics[width=0.45\textwidth]{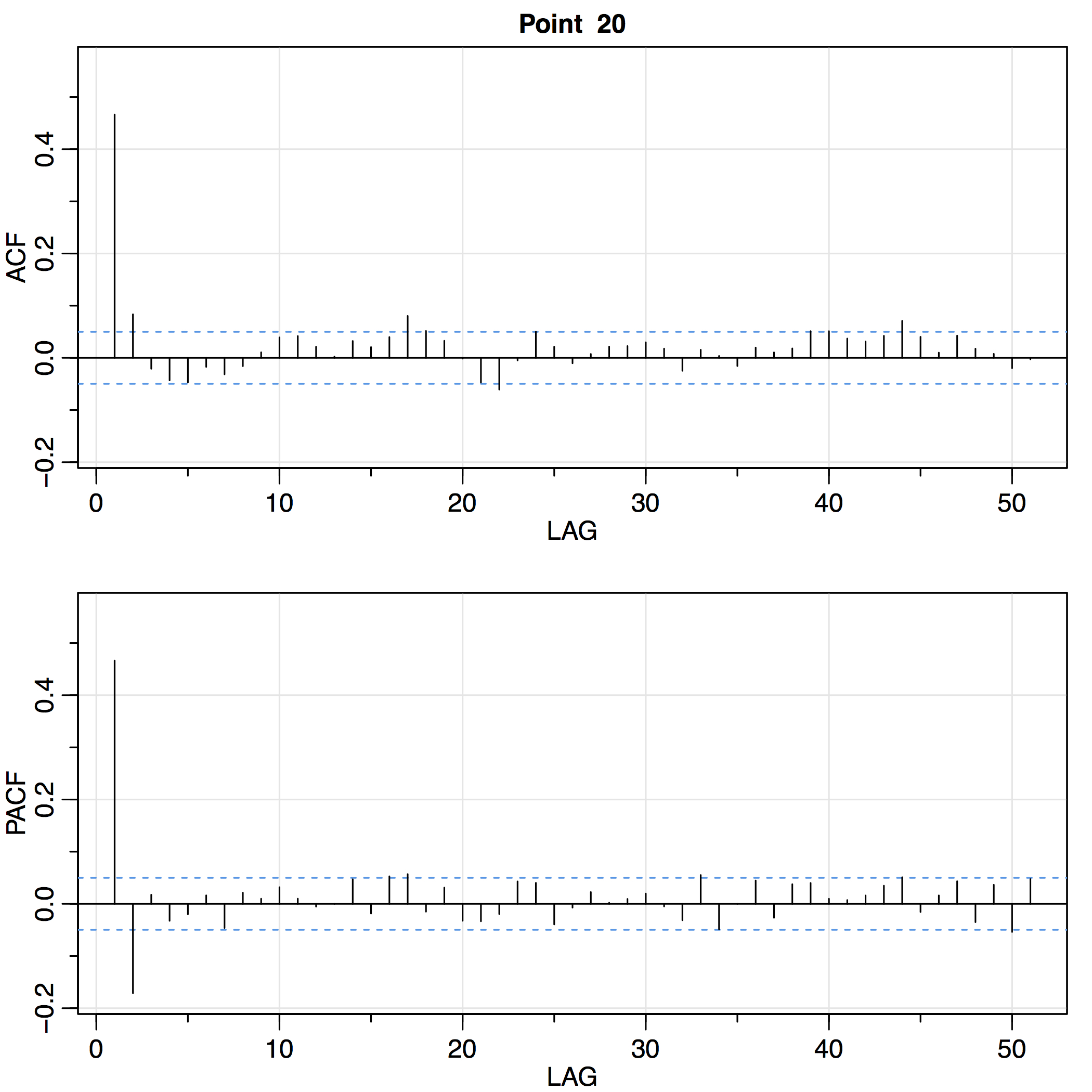}} 
      \subfigure[]{\includegraphics[width=0.45\textwidth]{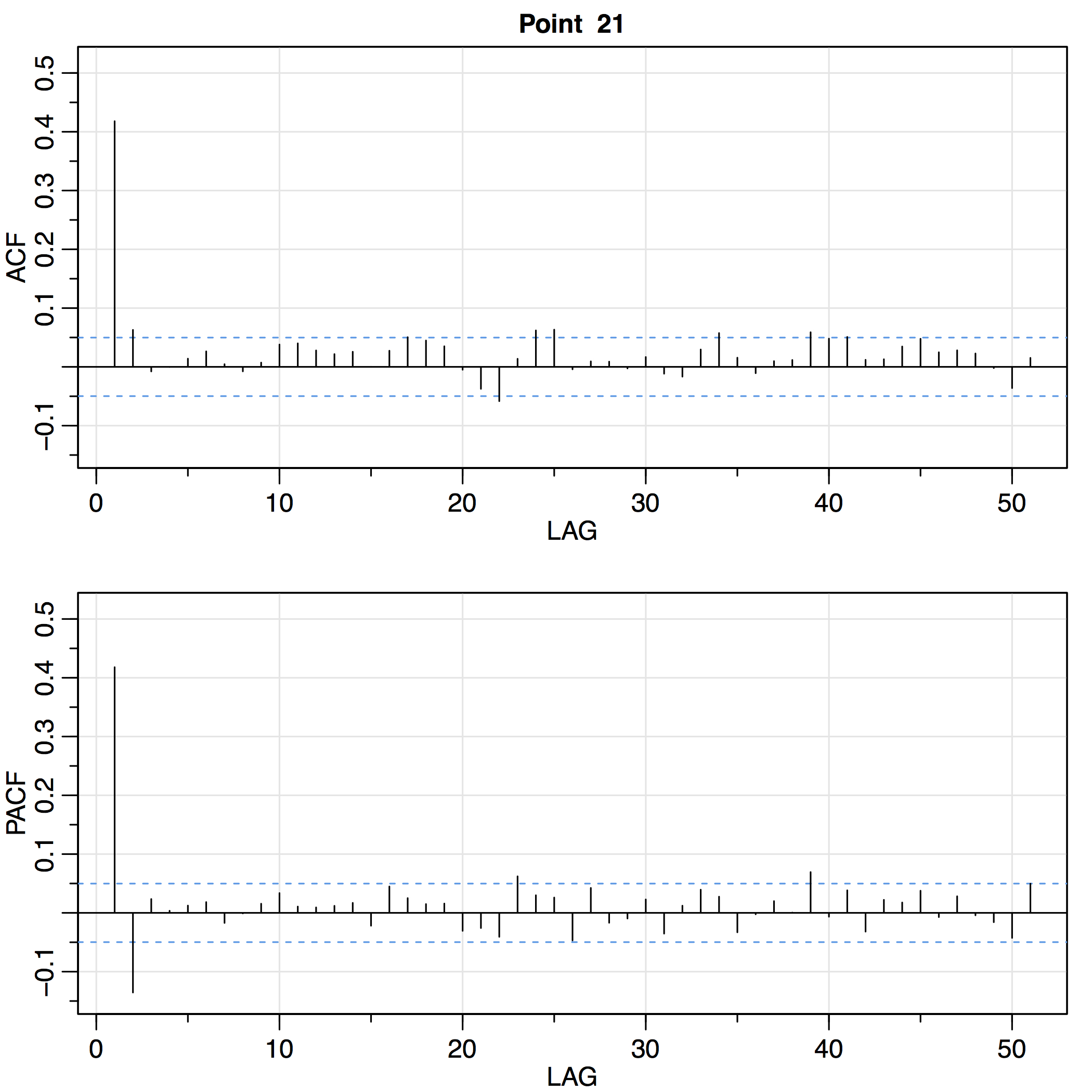}} 
    \subfigure[]{\includegraphics[width=0.45\textwidth]{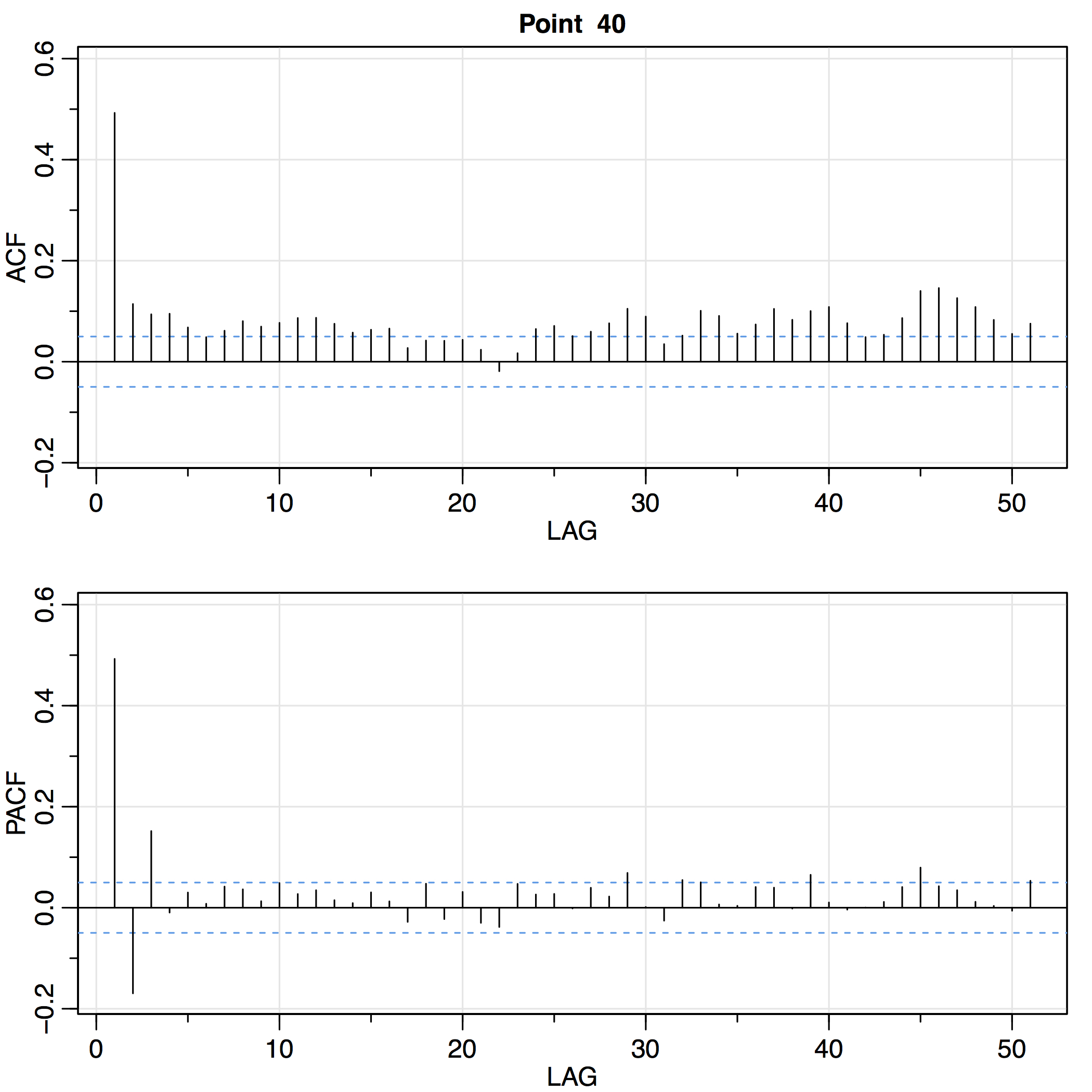}} 
      \subfigure[]{\includegraphics[width=0.45\textwidth]{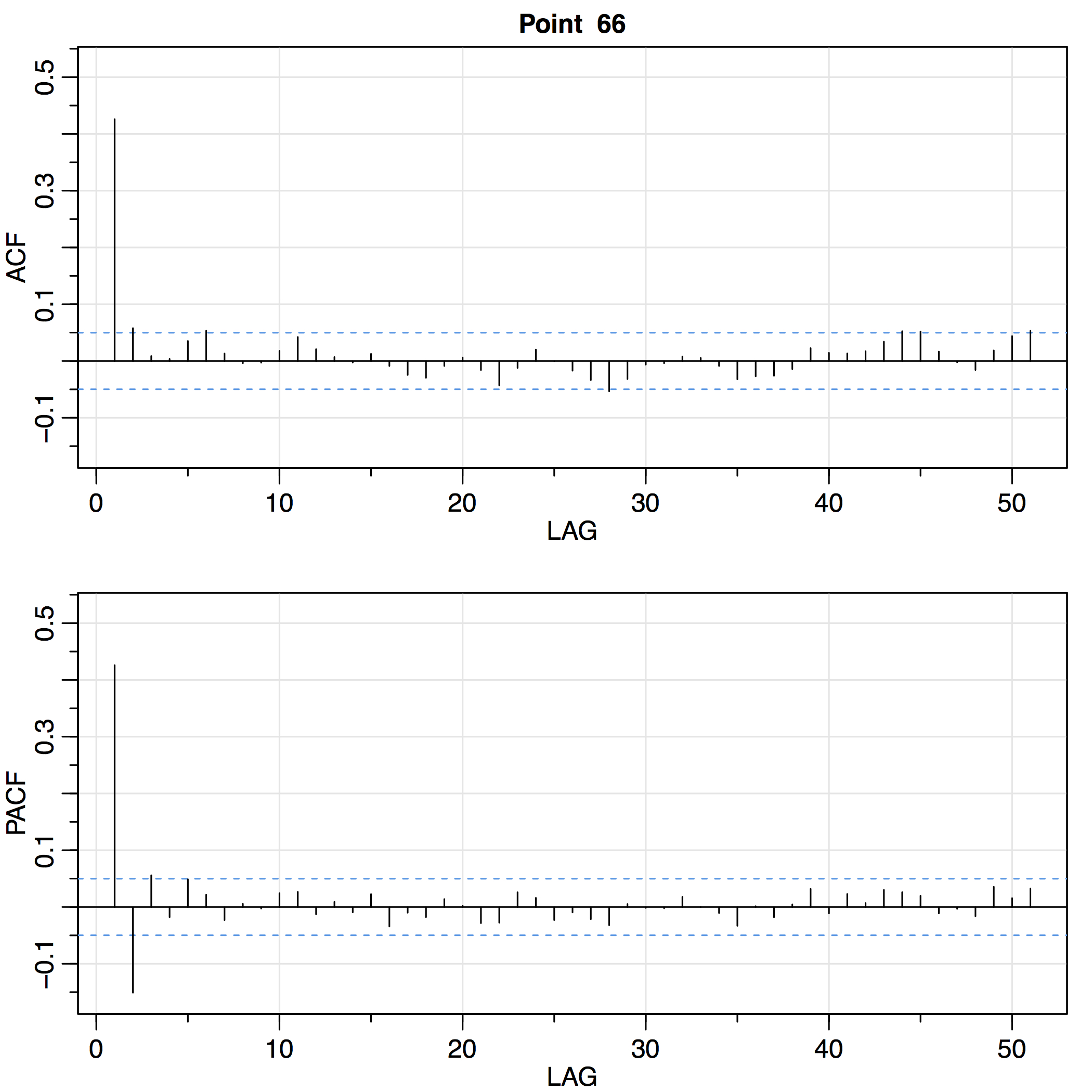}} 
  
 \caption{Autocorrelation and partial autocorrelation functions for deseasonalized wind speeds in the Mojave Desert (a,b), Coachella Valley (c), Mojave Valley (d)}
  \label{acf}
\end{figure}

The deseasonalized data exhibit significant autocorrelation effects (Fig~\ref{acf}) that we choose to address by an AR(2) model, as expressed in Eq~\ref{ar(2)}. Indeed, a lag of 2 days is common for nearly half of the 85 time series, including in the desert area of Mojave. We acknowledge that there are points on the grid where a higher order AR process can be more appropriate. For instance, the partial autocorrelation function in the time series of Point 40 (Fig~\ref{acf} (c)) suggests an AR(3) model. In few other locations, we observe higher autoregressive degrees as, for example, 5 or 7. Nevertheless, the improvement in terms of Akaike`s information criterion (AIC) when using a more complex AR model is between 0.07\% and 8.4\%. Thus, we decide to break the trade-off between complexity and accuracy by choosing an AR(2) model in all the 85 time series.

\begin{figure}[htbp]
 \centering
	 \subfigure[]{\includegraphics[width=0.4\textwidth]{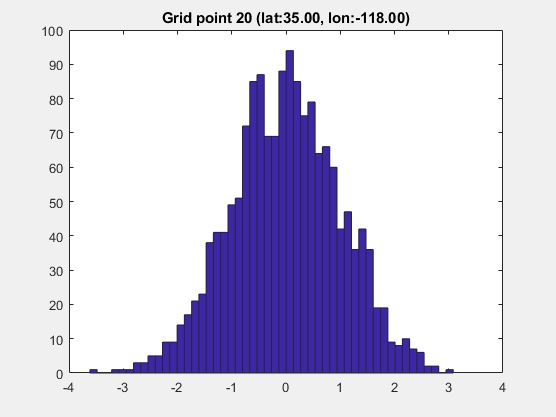}} 
	\subfigure[]{\includegraphics[width=0.3\textwidth]{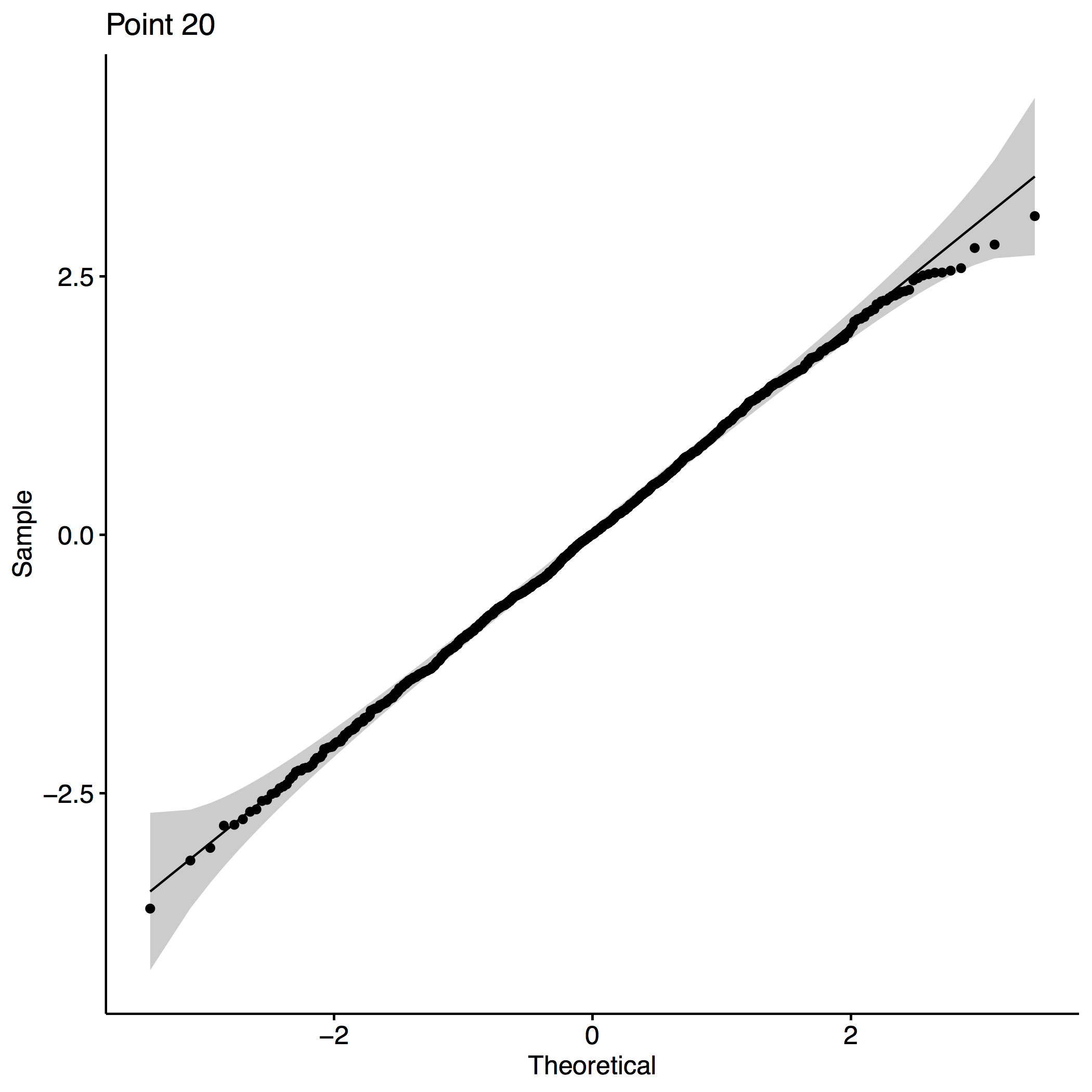}} 
	 \subfigure[]{\includegraphics[width=0.4\textwidth]{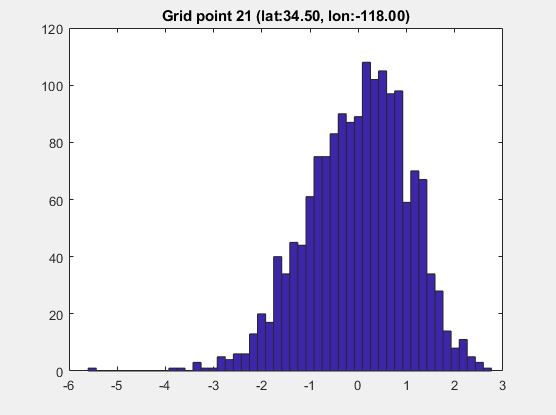}} 
	\subfigure[]{\includegraphics[width=0.3\textwidth]{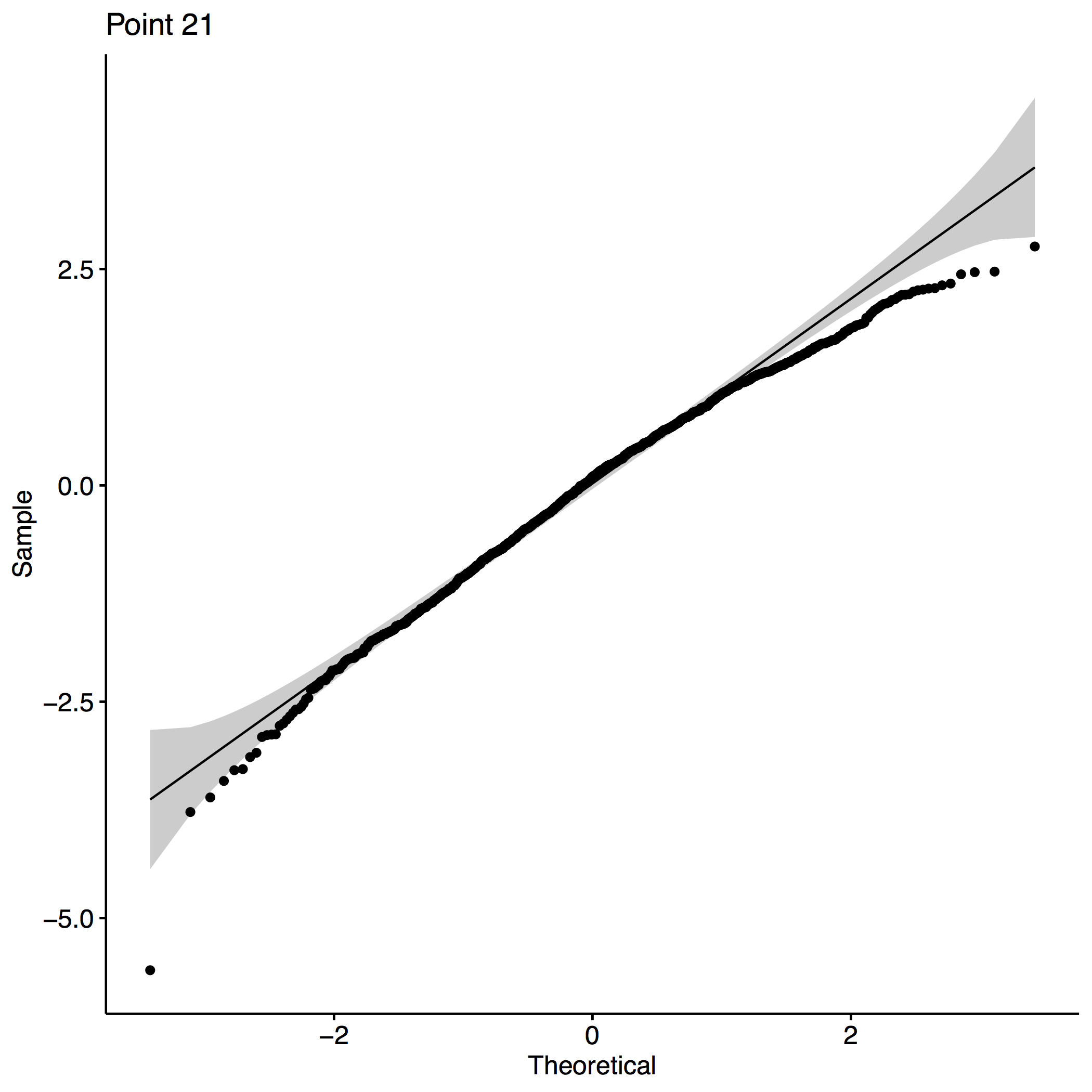}}
	 \subfigure[]{\includegraphics[width=0.4\textwidth]{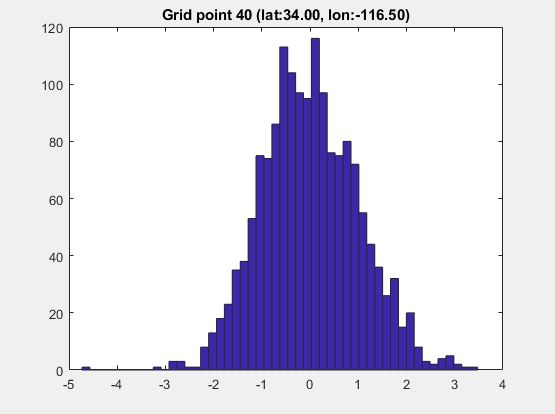}} 
 	\subfigure[]{\includegraphics[width=0.3\textwidth]{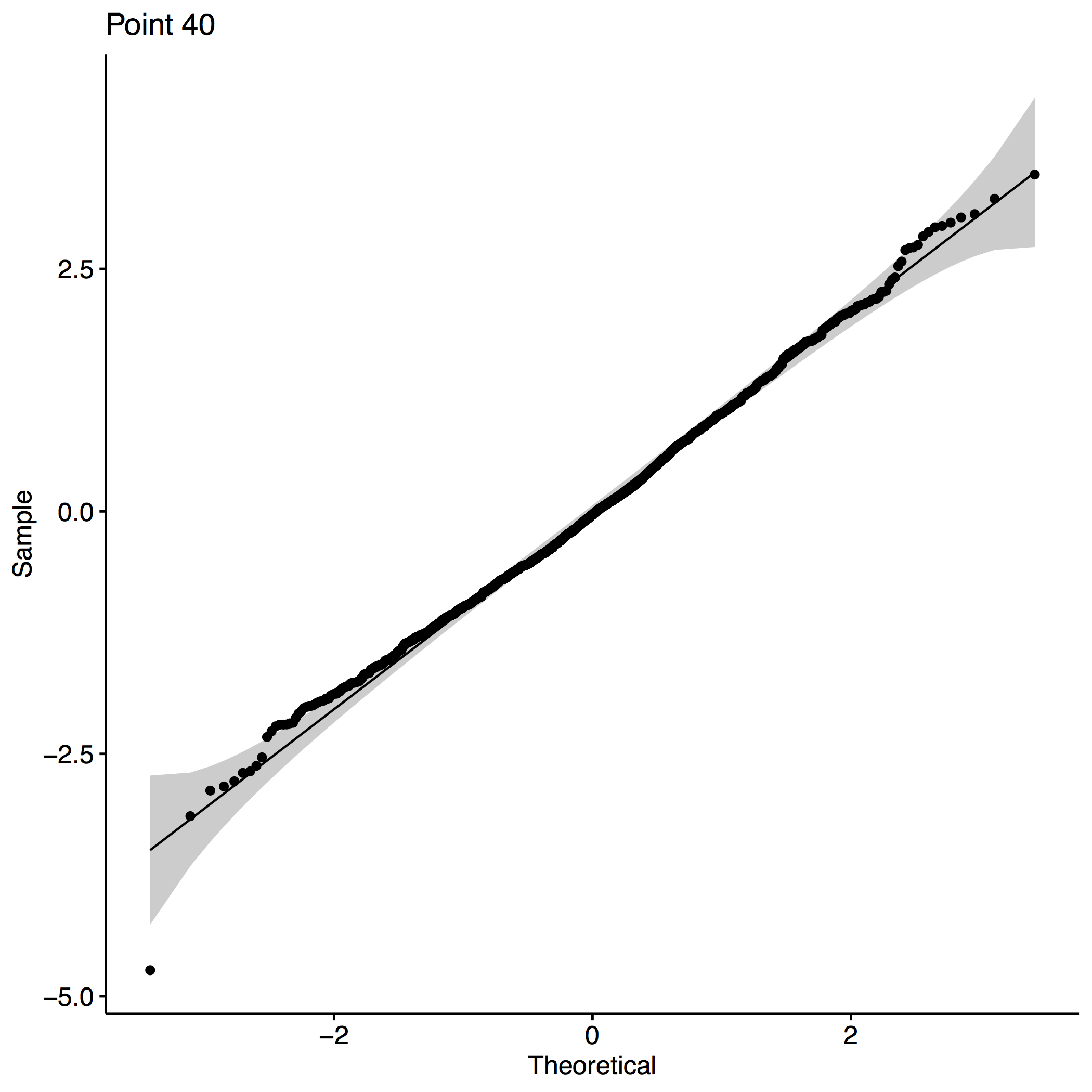}} 
	  \subfigure[]{\includegraphics[width=0.4\textwidth]{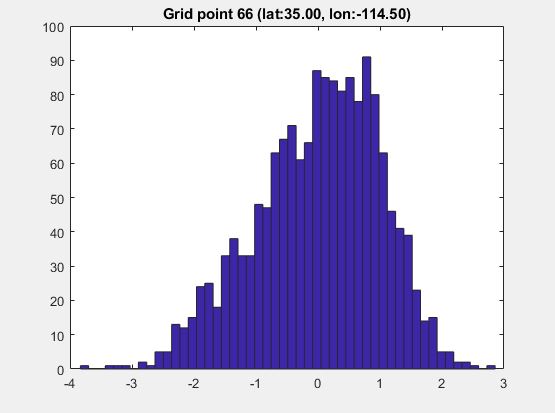}} 
	\subfigure[]{\includegraphics[width=0.3\textwidth]{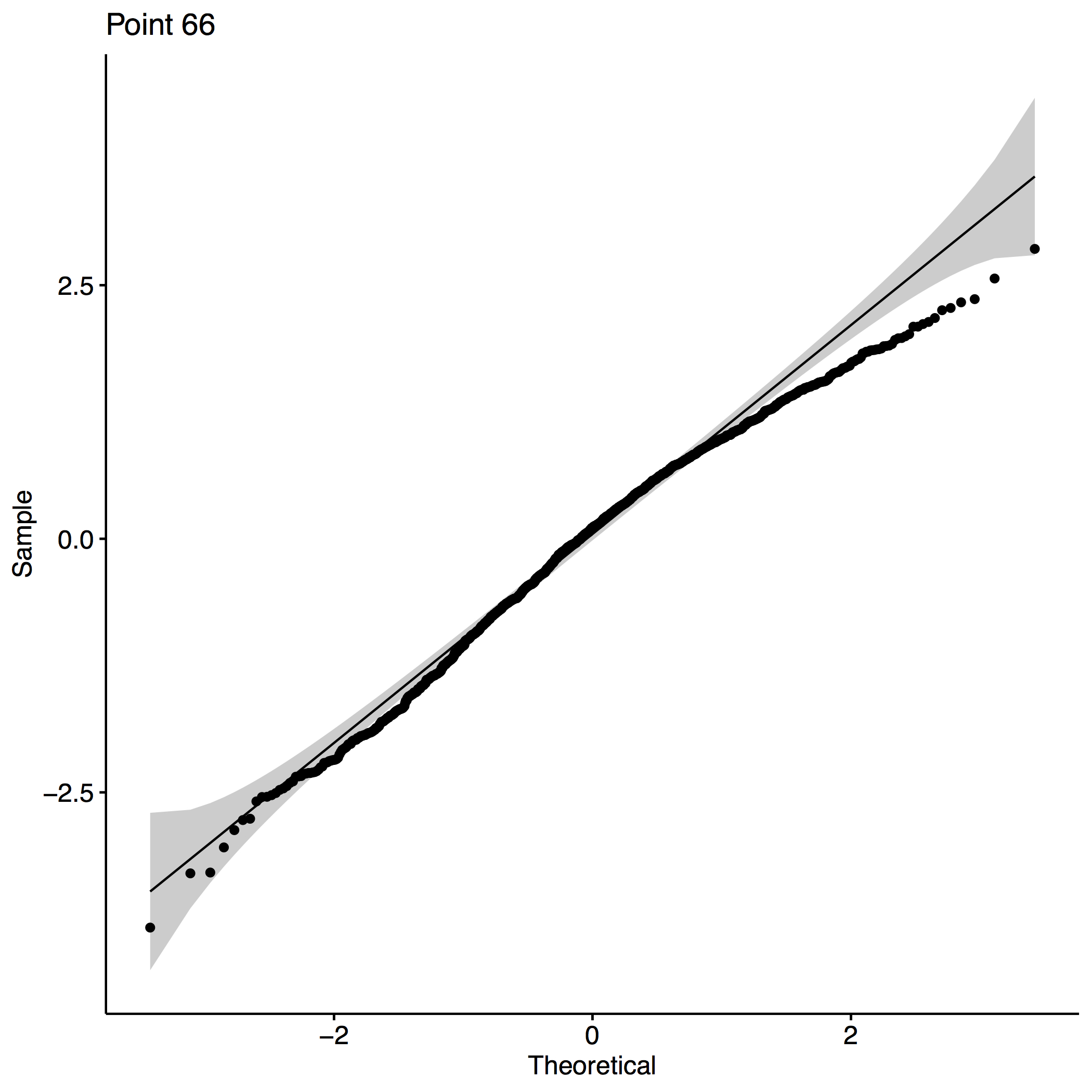}}
 \caption{Histograms and Q-Q plots of the final residuals in the Mojave Desert (a, b, c, d), Coachella Valley (e, f), Mojave Valley (g, h)}
  \label{normalresiduals}
\end{figure}

Moreover, we approach the heteroskedasticity in the residuals by fitting a Fourier series truncated to 1 term as seen from Eq~\ref{epsilon}. The final residuals $\epsilon(s_k,t), k=1,\dots 85$ are assumed to be normally distributed in our theoretical model. Fig~\ref{normalresiduals} gives a visual account on normality via histograms and Q-Q plots. We also performed a Kolmogorov-Smirnov normality test at the 5\% significance level. The Q-Q plots suggest that it is reasonable to assume normality in all the 85 model residuals. And the normality assumption through the statistical test is verified in 51 out of the 85 residuals. We note that both the visual and the statistical methods showed that the residuals in Point 20 in the Mojave Desert and in Point 40 in Coachella Valley are normally distributed. However, we observe less agreement between these methods in certain scattered sites in the Mojave Desert and westward. We believe this is due to the complexity of the local topography (mountainous area surrounding the desert valley) and to other effects undetectable on a daily wind speed scale (wind gusts).

All the parameters for $W(s_k,t), k=1,\dots 85$ are depicted in Table~\ref{allparams}.

Next, we validate our model in 4 grid points for the out-of-sample period between 1 July 2019 and 1 March 2020. In all the different sites, our time model provides better results than the persistence model\footnote{The trivial model that assumes the same wind speed values tomorrow as today}. In terms of mean absolute percentage error (MAPE), our model performs better than persistence by 5.7-18.8\% in the Mojave Desert area (Points 20, 21), 3.5\% in San Jacinto Mountains north of Palm Springs (Point 40) and 10.9\% along the Colorado River (Point 66).

\subsection{Spatial model}

\begin{figure}[htbp]
  \centering
    \subfigure[]{\includegraphics[width=0.45\textwidth]{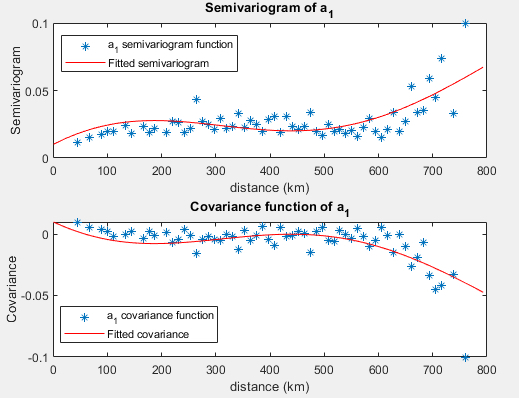}} 
      \subfigure[]{\includegraphics[width=0.437\textwidth]{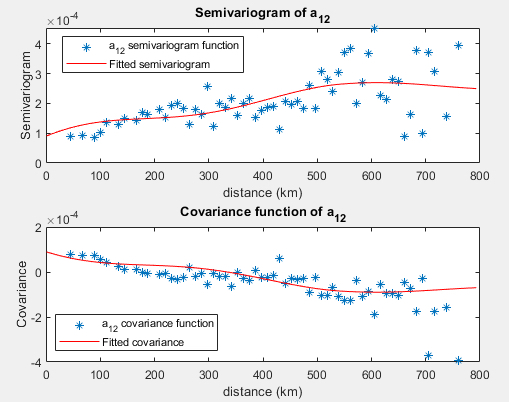}} 
    \subfigure[]{\includegraphics[width=0.45\textwidth]{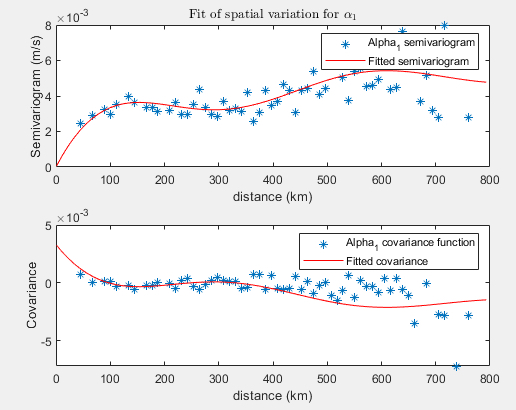}} 
      \subfigure[]{\includegraphics[width=0.45\textwidth]{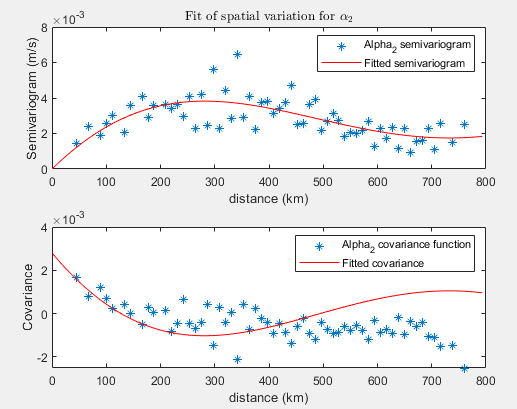}} 
     \subfigure[]{\includegraphics[width=0.45\textwidth]{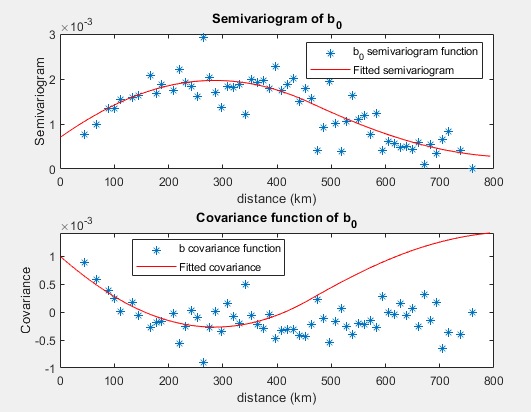}} 
 	\subfigure[]{\includegraphics[width=0.43\textwidth]{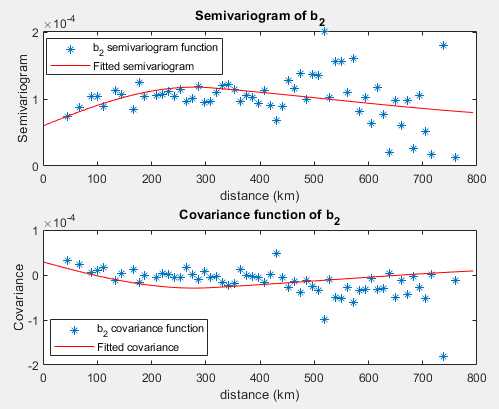}}

 \caption{Fitted semivariograms and covariances for $a_1,a_{12}, \alpha_1,\alpha_2,b_0,b_2$}
  \label{fittedvariograms}
\end{figure}

The spatial variability of each estimated temporal parameter spans over nearly 200000 km$^2$. This offers a wealth of statistical information that we quantify in semivariograms as in Fig~\ref{fittedvariograms}. 

Kriging on such a wide space as Southern California has the additional benefit of error reduction due to spatial smoothing. This hinges on the fact that the estimator for $W(s_0,t), \forall s_0\in D^\prime$ given in Eq~\ref{krigingestimator} is in fact a non-exact interpolator (\cite{Cressie}). In this respect, the size of the region and the number of sites contained have been proven by \cite{Fockenetal} to decide the error reduction.

\begin{figure}[htbp]
  \centering
    \subfigure[]{\includegraphics[width=0.45\textwidth]{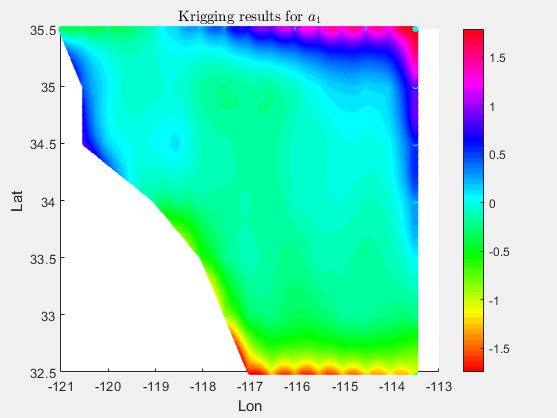}} 
      \subfigure[]{\includegraphics[width=0.45\textwidth]{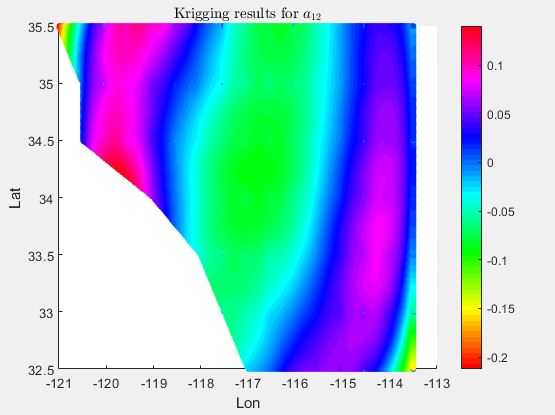}} 
    \subfigure[]{\includegraphics[width=0.45\textwidth]{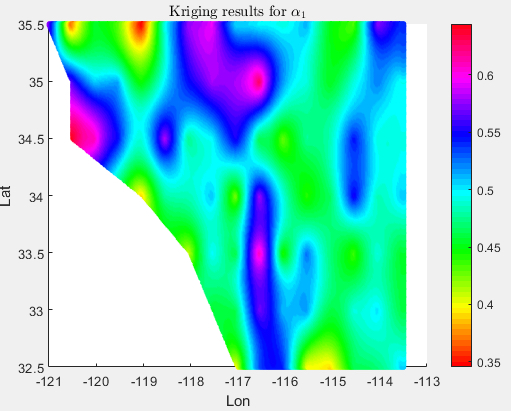}} 
      \subfigure[]{\includegraphics[width=0.45\textwidth]{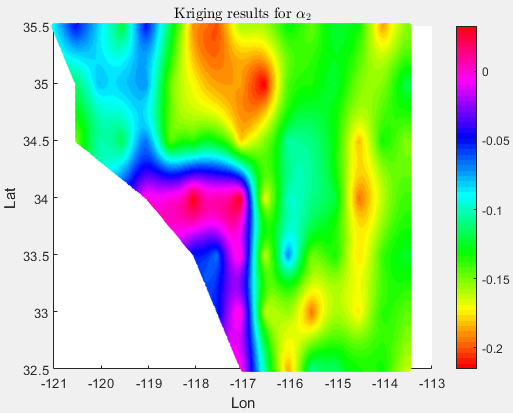}} 
     \subfigure[]{\includegraphics[width=0.45\textwidth]{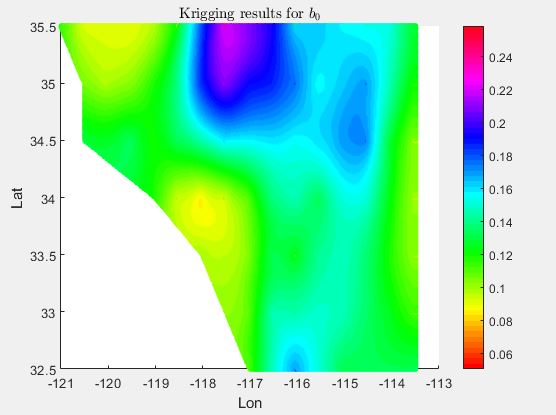}} 
 	\subfigure[]{\includegraphics[width=0.45\textwidth]{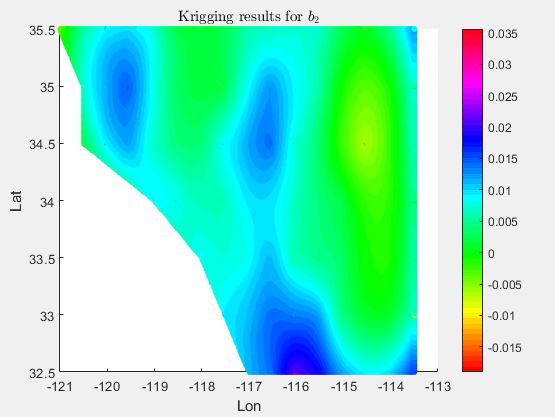}} 

 \caption{Kriging predictor maps for $a_1,a_{12}, \alpha_1,\alpha_2,b_0,b_2$}
  \label{krigingmaps}
\end{figure}

In the present study we perform ordinary kriging on a discretized space of 0.01\degree latitude and longitude. In other words, we choose successively $s_0$ in the optimization problem (Eq~\ref{optimizationproblem}) only 0.01\degree apart from the other known locations ($s_1,s_2 \dots s_{85}$). We do this for all the parameters in our temporal model. Fig~\ref{krigingmaps} shows our kriging predictor maps for $a_1$, $a_{12}$, $\alpha_1$, $\alpha_2$, $b_0$, $b_2$ after we have solved the optimization problem in Eq~\ref{krigingestimator} a considerable amount of times.

We can infer on certain aspects by superposing our resulting maps onto the geographic map of Southern California. Seasonal effects tend to be stronger along the upper coast or along the Colorado River (rightmost side of our grid). Maps of the other seasonality terms not presented in the paper show clear cyclic effects particular to the Mojave Desert area. The predictor maps of the autoregressive terms reflect rather accurately the particularities of the Californian landscape. The variations in the Mojave Desert area are rather clear. Otherwise, the flat areas seem strongly correlated on a 1-day lag. Wind speeds in the highly populated area along the coast from Los Angeles to San Diego seem nearly uncorrelated on a 2-day lag. It is noteworthy that the bias in our seasonal variance (i.e. $b_0$ term) is higher in the Mojave Desert than elsewhere.

\subsection{Spatio-temporal model: benchmarking results}

Upon fitting our model in time and space, we proceed to finding in-sample and out-of-sample one day ahead wind speed estimates at METAR locations. We summarize our results in terms of the first 4 standard moments of the errors and amount of estimates outside the 95\% prediction interval (PI). In Table~\ref{benchmark} we detail these measures for 7 METAR locations.

\newpage
\begin{center}
\begin{longtable}{c  c c c c c}

\caption{Benchmark of spatio-temporal model}\\
\hline
\textbf{Airport} & \textbf{Mean} & \textbf{Std} & \textbf{Skewness} & \textbf{Kurtosis} & \textbf{\% outside PI} \\
\hline
\textbf{} & \textbf{} & \textbf{} & \textbf{In-sample} & \textbf{} & \textbf{}\\

\hline
KEDW&-0.2103&0.4887&0.4326&3.3309&4.8507\\
KWJF&-0.304&0.4712&0.3769&2.8486&5.7728\\
KPMD&-0.2855&0.3969&0.3203&3.0118&4.1563\\
KSDB&-0.4793&0.2733&0.1719&3.3660&5.1745\\
KPSP&0.0014&0.3254&0.0196&4.1450&1.7391\\
KNJK&-0.0056&0.4391&-0.6256&3.7759&3.1815\\
KEED&-0.0237&0.3954&0.2176&3.3842&2.7044\\
\hline
\textbf{} & \textbf{} & \textbf{} & \textbf{Out-of-sample} & \textbf{} & \textbf{}\\
\hline
KEDW&-0.1090&0.5755&0.3245&2.7142&4.8979\\
KWJF&-0.2136&0.5100&0.1371&2.4045&4.0816\\
KPMD&-0.2059&0.4268&0.3077&2.7283&3.2921\\
KSDB&-0.4931&0.2964&-0.1946&2.8488&4.6135\\
KPSP&0.0416&0.3186&0.0901&3.9657&1.2244\\
KNJK&-0.0358&0.4113&-0.9451&4.4134&2.0408\\
KEED&-0.0027&0.4062&-0.1036&3.2713&2.9535\\
\hline
\label{benchmark}
\end{longtable}
\end{center}

We obtain the best fit at Palm Springs Airport (KPSP) in the valley of Jacinto Mountain. The errors are nearly symetrically distributed with respect to their zero mean while very few observations are scattered outside the 95\%PI.

The first 4 airports in our list are in the proximity of the Mojave Desert. In this area, we have the highest percentage of observations outside the 95\% PI. However, in the time span of the out-of-sample data, none of the sites have more than 5\% measurements beyond the PI.

The negative mean of errors suggest an almost ubiquitous bias between our estimates and METAR observations. Indeed, this seems to be a consequence of the fact that the former are estimates at 100 m AGL while the latter are measurements from 10 m AGL up-scaled by a simplified model (Eq~\ref{loglaw}). This bias can potentially be corrected by adjusting a different roughness length coefficient $z_0$.

\section{Conclusions}

This article proposes a spatio-temporal model for day-ahead predictions of wind speed in any location of interest. Our method employs AR techniques in time and ordinary kriging on parameters for the space domain.  We construct this model in three steps on a widespread grid of data from Southern California.

First, we model the time series in each grid point $\{ W(s_k,t) | k=1,2\dots 85 \}$ in the style of \cite{BenthBook}, \cite{SaltyteBenthBenth}, \cite{SaltyteBenthSaltyte}. We do this in stages by considering a seasonal, an aturoregressive and a seasonal variance part. In total, we fit 17 parameters for 85 different time series. Out-of-sample results from 4 grid points show that this performs up to 18\% better than the persistence model.

Second, assuming intrinsic stationarity, we study the temporal parameters derived earlier for each grid point. We determine the spatial variations of each parameter with the aid of semivariograms. We infer on their theoretical models and use them to determine ordinary kriging estimators for our parameters (\cite{CressieWikle}, \cite{Cressie}). They are optimal unbiased estimators of the type: 
$$\hat{a}_i(s_0)=\sum_{k=1}^{85}\lambda_k a_i(s_k), $$ 
with $a_i$ being any of the 17 temporal parameters derived earlier. We obtain 0.01\degree dense maps with temporal parameters. Their spatial variation is meaningful in the geographical context.

Finally, we choose any new location whereby we want to predict day ahead wind speeds. In our case study, we choose those sites where we have METAR observations for as long as our time sample goes. By doing so, we take our model out-of-sample in the time and space realms. Results show that all the sites have less than 5\% measurements outside the 95\% PI. We conclude that our spatio-temporal model performs well given the particularities of the Southern California landscape.

By the techniques and the data employed, our work provides a hybrid spatio-temporal model combining physical, conventional and spatial statistical methods. In practice, this performs well on a complex terrain as shown in our case study. This research is our attempt to respond to the academic and market need for accurate wind speed forecasting. 

We see potential improvements in our methodology from several vantage points. On the one hand, the model could be fine-tuned for hourly data. Perhaps the most immediate way to do this would be by splitting our data in 24 daily time series. Also, more sophisticated temporal models could be employed. For example, one could fit an ARIMA model of different dimensions in each point and a GARCH model for residuals. On the other hand, other pieces of information about the terrain or weather state can be incorporated (altitude, air temperature, humidity etc.). The roughness length coefficient $z_0$ could be further adjusted to avoid the 10 to 100 m AGL bias explained in Section~\ref{space-time problem}. We neglected wind direction in this study. Including this information in the spatial semivariogram could refine the results.

\section*{Acknowledgement}
Mihaela Puica is grateful to Refinitiv Commodities Content \& Research for providing us with the static data underlying this study. 
Fred Espen Benth acknowledges financial support from SPATUS, a project funded by UiO: Energy (University of Oslo). 

The authors declare that there is no conflict of interest.

\newpage

\bibliography{articlebib_Puica_Benth}

\begin{thebibliography}{10}
\expandafter\ifx\csname url\endcsname\relax
  \def\url#1{\texttt{#1}}\fi
\expandafter\ifx\csname urlprefix\endcsname\relax\def\urlprefix{URL }\fi
\expandafter\ifx\csname href\endcsname\relax
  \def\href#1#2{#2} \def\path#1{#1}\fi

\bibitem{IRENAReport}
IRENA, Deployment, investment, technology, grid integration and socio-economic
  aspects (a global energy transformation paper), ISBN 978-92-9260-155-3
  (2019).

\bibitem{LangeFocken}
M.~{Lange}, U.~{Focken}, New developments in wind energy forecasting, in: 2008
  IEEE Power and Energy Society General Meeting - Conversion and Delivery of
  Electrical Energy in the 21st Century, 2008, pp. 1--8.
\newblock \href {http://dx.doi.org/10.1109/PES.2008.4596135}
  {\path{doi:10.1109/PES.2008.4596135}}.

\bibitem{Hodgeetal}
B.-M. {Hodge}, A.~{Florita}, J.~{Sharp}, M.~{Margulis}, D.~{Mcreavy}, The value
  of improved short- term wind power forecasting, Tech. Rep.
  NREL/TP-5D00-63175, National Renewable Energy Laboratory (NREL) (2015).

\bibitem{Leietal}
M.~Lei, L.~Shiyan, J.~Chuanwen, L.~Hongling, Z.~Yan, A review on the
  forecasting of wind speed and generated power, Renewable and Sustainable
  Energy Reviews 13~(4) (2009) 915 -- 920.
\newblock \href {http://dx.doi.org/https://doi.org/10.1016/j.rser.2008.02.002}
  {\path{doi:https://doi.org/10.1016/j.rser.2008.02.002}}.

\bibitem{Botterudetal}
A.~Botterud, J.~Wang, V.~Miranda, R.~J. Bessa, Wind power forecasting in u.s.
  electricity markets, The Electricity Journal 23~(3) (2010) 71 -- 82.
\newblock \href {http://dx.doi.org/https://doi.org/10.1016/j.tej.2010.03.006}
  {\path{doi:https://doi.org/10.1016/j.tej.2010.03.006}}.

\bibitem{Costaetal}
A.~Costa, A.~Crespo, J.~Navarro, G.~Lizcano, H.~Madsen, E.~Feitosa, A review on
  the young history of the wind power short-term prediction, Renewable and
  Sustainable Energy Reviews 12~(6) (2008) 1725--1744.
\newblock \href {http://dx.doi.org/10.1016/j.rser.2007.01.015}
  {\path{doi:10.1016/j.rser.2007.01.015}}.

\bibitem{Chang}
W.-Y. Chang, A literature review of wind forecasting methods, Journal of Power
  and Energy Engineering 2 (2014) 161--168.
\newblock \href {http://dx.doi.org/110.4236/jpee.2014.24023}
  {\path{doi:110.4236/jpee.2014.24023}}.

\bibitem{BenthBook}
F.E.{Benth}, J.S.{Benth}, Modeling and Pricing in Financial Markets for Weather
  Derivatives, Advanced series on statistical science \& applied probability,
  World Scientific, 2013.

\bibitem{SaltyteBenthBenth}
J.~\v{S}altyt\.{e} Benth, F.~E. Benth, Analysis and modelling of wind speed in
  new york, Journal of Applied Statistics 37~(6) (2010) 893--909.
\newblock \href {http://dx.doi.org/10.1080/02664760902914490}
  {\path{doi:10.1080/02664760902914490}}.

\bibitem{Giebeletal}
G.~Giebel, R.~Brownsword, G.~Kariniotakis, M.~Denhard, C.~Draxl, The
  State-Of-The-Art in Short-Term Prediction of Wind Power: A Literature
  Overview, 2nd edition, ANEMOS.plus, 2011, project funded by the European
  Commission under the 6th Framework Program, Priority 6.1: Sustainable Energy
  Systems.
\newblock \href {http://dx.doi.org/10.11581/DTU:00000017}
  {\path{doi:10.11581/DTU:00000017}}.

\bibitem{Lebassi-Habtezionetal}
B.~Lebassi-Habtezion, J.~González, R.~Bornstein, Modeled large-scale warming
  impacts on summer california coastal-cooling trends, Journal of Geophysical
  Research: Atmospheres 116~(D20).
\newblock \href {http://dx.doi.org/https://doi.org/10.1029/2011JD015759}
  {\path{doi:https://doi.org/10.1029/2011JD015759}}.

\bibitem{Loetal}
J.~C.-F. Lo, Z.-L. Yang, R.~A. Pielke~Sr., Assessment of three dynamical
  climate downscaling methods using the weather research and forecasting (wrf)
  model, Journal of Geophysical Research: Atmospheres 113~(D9).
\newblock \href {http://dx.doi.org/https://doi.org/10.1029/2007JD009216}
  {\path{doi:https://doi.org/10.1029/2007JD009216}}.

\bibitem{Gutierrezetal}
J.~M. Gutierrez, A.~S. Cofino, R.~Cano, M.~A. Rodriguez, Clustering methods for
  statistical downscaling in short-range weather forecasts, Monthly Weather
  Review 132~(9) (01 Sep. 2004) 2169 -- 2183.
\newblock \href
  {http://dx.doi.org/10.1175/1520-0493(2004)132<2169:CMFSDI>2.0.CO;2}
  {\path{doi:10.1175/1520-0493(2004)132<2169:CMFSDI>2.0.CO;2}}.

\bibitem{Pryoretal}
S.~C. Pryor, J.~T. Schoof, R.~J. Barthelmie, Empirical downscaling of wind
  speed probability distributions, Journal of Geophysical Research: Atmospheres
  110~(D19).
\newblock \href {http://dx.doi.org/https://doi.org/10.1029/2005JD005899}
  {\path{doi:https://doi.org/10.1029/2005JD005899}}.

\bibitem{CassolaBurlando}
F.~Cassola, M.~Burlando, Wind speed and wind energy forecast through kalman
  filtering of numerical weather prediction model output, Applied Energy 99
  (2012) 154 -- 166.
\newblock \href
  {http://dx.doi.org/https://doi.org/10.1016/j.apenergy.2012.03.054}
  {\path{doi:https://doi.org/10.1016/j.apenergy.2012.03.054}}.

\bibitem{Sailoretal}
D.~Sailor, T.~Hu, X.~Li, J.~Rosen, A neural network approach to local
  downscaling of gcm output for assessing wind power implications of climate
  change, Renewable Energy 19~(3) (2000) 359 -- 378.
\newblock \href
  {http://dx.doi.org/https://doi.org/10.1016/S0960-1481(99)00056-7}
  {\path{doi:https://doi.org/10.1016/S0960-1481(99)00056-7}}.

\bibitem{Michelangelietal}
P.-A. Michelangeli, M.~Vrac, H.~Loukos, Probabilistic downscaling approaches:
  Application to wind cumulative distribution functions, Geophysical Research
  Letters 36~(11).
\newblock \href {http://dx.doi.org/https://doi.org/10.1029/2009GL038401}
  {\path{doi:https://doi.org/10.1029/2009GL038401}}.

\bibitem{TangBassill}
B.~H. Tang, N.~P. Bassill, Point downscaling of surface wind speed for forecast
  applications, Journal of Applied Meteorology and Climatology 57~(3) (01 Mar.
  2018) 659 -- 674.
\newblock \href {http://dx.doi.org/10.1175/JAMC-D-17-0144.1}
  {\path{doi:10.1175/JAMC-D-17-0144.1}}.

\bibitem{HaslettRaftery}
J.~Haslett, A.~E. Raftery, Space-time modelling with long-memory dependence:
  Assessing ireland's wind power resource, Journal of the Royal Statistical
  Society. Series C (Applied Statistics) 38~(1) (1989) 1--50.

\bibitem{Akylasetal}
E.{Akylas}, M.{Tombrou}, J.{Panourgias}, D.{Lalas}, The use of common
  meteorological predictions in estimating short term wind energy production in
  complex terrain, in: Proc. EWEC’97, 6 – 9 October, Dublin, Ireland, 1997.

\bibitem{Beyeretal}
H.~{Beyer}, A.~{Luig}, D.~{Heinemann}, U.~{Focken}, M.~{Lange}, E.~{Lorenz},
  B.~{Lueckehe}, K.~{Moennich}, H.~{Waldl}, Short term predictions for the
  power output of ensembles of wind turbines and pv-generators (Jul 2000).

\bibitem{PalomaresCastro}
A.{Palomares}, M.{de Castro}, Short-term wind prediction model at the strait of
  gibraltar based on a perfect prognosis statistical downscaling method, in:
  Proceedings of European wind energy conference, Madrid, 2003, 2003.

\bibitem{Nfaoui}
H.~Nfaoui, J.~Buret, A.~Sayigh, Stochastic simulation of hourly average wind
  speed sequences in tangiers (morocco), Solar Energy 56~(3) (1996) 301 -- 314.
\newblock \href
  {http://dx.doi.org/https://doi.org/10.1016/0038-092X(95)00103-X}
  {\path{doi:https://doi.org/10.1016/0038-092X(95)00103-X}}.

\bibitem{ErdemShi}
E.~Erdem, J.~Shi, Arma based approaches for forecasting the tuple of wind speed
  and direction, Applied Energy 88~(4) (2011) 1405 -- 1414.
\newblock \href
  {http://dx.doi.org/https://doi.org/10.1016/j.apenergy.2010.10.031}
  {\path{doi:https://doi.org/10.1016/j.apenergy.2010.10.031}}.

\bibitem{Damousisetal}
I.~G. {Damousis}, M.~C. {Alexiadis}, J.~B. {Theocharis}, P.~S. {Dokopoulos}, A
  fuzzy model for wind speed prediction and power generation in wind parks
  using spatial correlation, IEEE Transactions on Energy Conversion 19~(2)
  (2004) 352--361.
\newblock \href {http://dx.doi.org/10.1109/TEC.2003.821865}
  {\path{doi:10.1109/TEC.2003.821865}}.

\bibitem{Corotisetal}
R.~B. Corotis, A.~B. Sigl, M.~P. Cohen, Variance analysis of wind
  characteristics for energy conversion, Journal of Applied Meteorology and
  Climatology 16~(11) (01 Nov. 1977) 1149--1157.
\newblock \href
  {http://dx.doi.org/10.1175/1520-0450(1977)016<1149:VAOWCF>2.0.CO;2}
  {\path{doi:10.1175/1520-0450(1977)016<1149:VAOWCF>2.0.CO;2}}.

\bibitem{Beyeretal1993}
H.~Beyer, J.~Luther, R.~Steinberger-Willms, Power fluctuations in spatially
  dispersed wind turbine systems, Solar Energy 50~(4) (1993) 297 -- 305.
\newblock \href
  {http://dx.doi.org/https://doi.org/10.1016/0038-092X(93)90025-J}
  {\path{doi:https://doi.org/10.1016/0038-092X(93)90025-J}}.

\bibitem{PalominoMartin}
I.~Palomino, F.~Mart?n, A simple method for spatial interpolation of the wind
  in complex terrain, Journal of Applied Meteorology and Climatology 34~(7) (01
  Jul. 1995) 1678--1693.
\newblock \href {http://dx.doi.org/10.1175/1520-0450-34.7.1678}
  {\path{doi:10.1175/1520-0450-34.7.1678}}.

\bibitem{Leeetal}
J.-T. Lee, H.-G. Kim, Y.-H. Kang, J.-Y. Kim, Determining the optimized hub
  height of wind turbine using the wind resource map of south korea, Journal of
  Applied Meteorology and Climatology 12~(15).
\newblock \href {http://dx.doi.org/https://doi.org/10.3390/en12152949}
  {\path{doi:https://doi.org/10.3390/en12152949}}.

\bibitem{SaltyteBenthSaltyte}
J.~\v{S}altyt\.{e} Benth, L.~\v{S}altyt\.{e}, Spatial-temporal model for wind
  speed in lithuania, Journal of Applied Statistics 38~(6) (2011) 1151--1168.
\newblock \href {http://dx.doi.org/10.1080/02664763.2010.491857}
  {\path{doi:10.1080/02664763.2010.491857}}.

\bibitem{CressieWikle}
N.~Cressie, C.~Wikle, Statistics for Spatio-Temporal Data, Wiley series in
  probability and statistics, Wiley, 2011.

\bibitem{Cressie}
N.~A.~C. Cressie, Statistics for spatio-temporal data, Wiley series in
  probability and statistics, Wiley, Hoboken, N.J, 2011.

\bibitem{Matheron}
G.~Matheron, Principles of geostatistics, Economic Geology 58~(8) (1963)
  1246--1266.
\newblock \href {http://dx.doi.org/10.2113/gsecongeo.58.8.1246}
  {\path{doi:10.2113/gsecongeo.58.8.1246}}.

\bibitem{CressieHuang}
N.~Cressie, H.-C. Huang, Classes of nonseparable, spatio-temporal stationary
  covariance functions, Journal of the American Statistical Association
  94~(448) (1999) 1330--1339.
\newblock \href {http://dx.doi.org/10.1080/01621459.1999.10473885}
  {\path{doi:10.1080/01621459.1999.10473885}}.

\bibitem{Fockenetal}
U.~Focken, M.~Lange, K.~Mönnich, H.-P. Waldl, H.~G. Beyer, A.~Luig, Short-term
  prediction of the aggregated power output of wind farms—a statistical
  analysis of the reduction of the prediction error by spatial smoothing
  effects, Journal of Wind Engineering and Industrial Aerodynamics 90~(3)
  (2002) 231 -- 246.
\newblock \href
  {http://dx.doi.org/https://doi.org/10.1016/S0167-6105(01)00222-7}
  {\path{doi:https://doi.org/10.1016/S0167-6105(01)00222-7}}.

\end{thebibliography}

\appendix
\newpage
\section{Descriptive statistics}\label{tablevalues}
\begin{center}
\begin{longtable}{c c c c c c c}

\caption{Descriptive statistics of the time series in the grid points}\\
\hline
\textbf{Point} & \textbf{Mean} & \textbf{Std} & \textbf{Skewness} & \textbf{Kurtosis} & \textbf{Min} &\textbf{Max}\\
\hline
\endfirsthead
\multicolumn{7}{c}%
{\tablename\ \thetable\ -- \textit{Continued from previous page}} \\
\hline
\textbf{Point} & \textbf{Mean} & \textbf{Std} & \textbf{Skewness} & \textbf{Kurtosis} & \textbf{Min} &\textbf{Max}\\
\hline
\endhead
\hline \multicolumn{7}{r}{\textit{Continued on next page}} \\
\endfoot
\hline
\endlastfoot
1 &4.50& 2.04&0.94&4.18&1.08&13.90\\
2&4.52&1.59&1.30&5.20&1.84&12.95\\
3&3.74&1.59&1.00&3.96&1.34&11.75\\
4&6.58&2.83&0.46&2.47&1.39&15.14\\
5&4.44&1.64&1.29&5.22&1.20&12.56\\
6&3.62&1.36&1.20&4.57&1.31&10.62\\
7&3.55&1.75&1.22&4.02&1.05&10.81\\
8&2.95&1.05&0.77&4.57&0.87&8.77\\
9&3.00&1.32&2.08&9.84&0.91&11.96\\
10&2.92&1.68&1.54&5.11&0.87&10.73\\
11&3.21&1.18&1.19&5.68&1.02&10.13\\
12&2.34&1.07&1.92&10.16&0.53&11.08\\
13&2.08&1.00&2.44&11.28&0.81&9.57\\
14&4.45&2.23&2.00&7.43&1.70&15.93\\
15&2.78&0.98&0.77&3.69&0.70&7.04\\
16&5.41&2.21&0.81&3.69&0.92&14.77\\
17&4.68&2.75&1.44&4.70&1.01&17.56\\
18&3.69&1.34&1.81&7.22&1.38&11.36\\
19&5.22&2.67&0.86&3.99&0.72&15.69\\
20&5.94&2.94&0.56&2.82&0.72&16.62\\
21&4.97&1.98&0.50&3.13&0.92&12.93\\
22&3.15&0.90&1.50&6.84&1.37&8.46\\
23&3.88&1.81&2.26&9.87&1.01&16.08\\
24&4.69&2.64&0.63&2.84&0.50&13.41\\
25&6.66&3.39&0.34&2.41&0.53&16.42\\
26&4.07&2.10&1.45&5.57&0.83&15.26\\
27&3.53&1.28&3.24&25.41&1.08&17.37\\
28&2.78&1.25&2.24&9.53&1.06&9.93\\
29&3.71&1.66&2.15&10.10&1.37&16.44\\
30&4.77&2.43&0.86&3.35&0.55&13.42\\
31&6.00&2.77&0.55&3.03&0.75&16.31\\
32&4.19&2.02&1.03&3.85&0.90&12.79\\
33&2.48&0.89&1.70&9.69&0.91&10.30\\
34&3.15&1.29&1.77&6.47&1.05&9.48\\
35&2.89&1.27&2.10&8.28&1.01&10.34\\
36&3.36&1.30&2.17&10.01&1.37&13.33\\
37&4.68&2.29&0.76&3.16&0.48&13.31\\
38&5.38&2.87&0.62&3.00&0.74&15.03\\
39&5.21&2.54&1.16&4.32&1.13&15.69\\
40&4.11&1.97&1.02&4.23&0.70&14.04\\
41&4.51&2.08&0.71&2.88&1.01&11.85\\
42&4.76&2.40&1.04&3.75&0.94&14.21\\
43&4.33&2.80&2.03&7.48&1.37&19.51\\
44&5.43&2.02&0.25&2.84&0.66&12.55\\
45&3.32&1.76&1.26&4.84&0.56&10.97\\
46&4.43&1.83&0.98&4.22&0.72&13.34\\
47&4.64&2.12&0.98&3.70&0.80&12.16\\
48&3.44&1.26&1.04&5.72&0.93&9.64\\
49&2.80&1.32&1.16&4.73&0.72&9.71\\
50&4.96&2.70&1.11&3.80&1.06&15.18\\
51&3.72&1.76&0.94&3.81&0.73&10.90\\
52&3.82&1.64&1.12&4.49&0.90&11.11\\
53&3.45&1.62&1.04&4.67&0.60&12.36\\
54&4.06&1.44&1.02&4.71&1.27&11.11\\
55&4.60&2.16&1.39&5.37&1.16&15.39\\
56&3.78&1.81&0.84&3.15&0.85&10.64\\
57&2.81&1.29&1.58&6.35&0.72&9.58\\
58&6.28&2.52&0.50&2.80&0.90&14.98\\
59&5.95&2.58&0.90&3.66&1.28&16.74\\
60&4.53&2.14&1.12&4.47&0.75&14.06\\
61&4.40&1.84&0.77&3.59&0.77&12.02\\
62&5.72&2.30&0.53&3.15&0.93&13.74\\
63&5.41&2.38&0.81&3.31&0.89&15.63\\
64&4.33&1.52&0.68&3.49&1.29&11.05\\
65&4.94&1.87&0.33&2.68&0.83&11.50\\
66&5.34&2.38&0.63&3.15&1.03&14.43\\
67&4.27&2.22&1.31&5.68&0.64&16.52\\
68&4.13&1.89&0.94&4.34&0.85&13.41\\
69&4.82&1.86&0.36&2.94&1.02&13.04\\
70&5.01&2.07&0.84&3.61&0.84&13.40\\
71&5.34&2.02&0.37&2.76&1.11&13.04\\
72&3.81&1.84&0.83&3.29&0.83&11.22\\
73&4.94&2.04&0.48&2.84&1.06&12.32\\
74&4.59&1.68&0.82&4.29&1.04&13.31\\
75&4.92&1.85&0.56&3.22&1.28&12.45\\
76&5.07&1.88&0.65&3.59&1.13&13.75\\
77&4.82&1.99&0.89&4.01&0.91&14.51\\
78&4.99&1.99&0.52&3.16&0.93&13.29\\
79&5.52&2.20&0.68&3.14&1.14&14.21\\
80&4.86&1.86&0.66&2.98&1.44&11.53\\
81&3.72&1.34&0.95&3.98&1.08&9.13\\
82&3.90&1.40&0.86&4.13&1.32&11.18\\
83&4.98&1.72&0.90&4.42&1.21&13.55\\
84&4.82&1.73&0.86&4.11&1.40&13.76\\
85&4.86&1.98&0.46&2.99&0.86&12.99
\label{gridpointsstatistics}
\end{longtable}
\end{center}

\newpage
\section{Ordinary Kriging equations}\label{derivation}
In Section~\ref{OK} we derived the ordinary kriging equations claiming that the objective function of the optimization problem from Eq~\ref{optimizationproblem} can be written conveniently.
\begin{proposition}
The mean-squared error of the fit $\hat{Y}(s_0)=\sum_{i=1}^{85}\lambda_i Y(s_0)$, with $\sum_{i=1}^{85}\lambda_i =1$ can be written as:
$$E \bigg[ (Y(s_0)-\hat{Y}(s_0))^2\bigg]  = - \sum_{i=1}^{85}\sum_{j=1}^{85}\lambda_i \lambda_j \gamma(s_i-s_j)+ 2\sum_{i=1}^{85}\gamma(s_i-s_0)$$
where $\gamma(\cdot)$ is the semivariogram of the process $Y$ 
\end{proposition}

\newenvironment{proof}{\paragraph{Proof:}}{\hfill$\square$}

\paragraph{Proof}:
By the definition of semivariogram, we have that:
\begin{align}
\gamma(s_i-s_j)&=\frac{1}{2} Var \bigg( Y(s_i)-Y(s_j)\bigg)=\frac{1}{2}E\bigg[ (Y(s_i)-Y(s_j))^2\bigg]= \nonumber \\
&=\frac{1}{2}E\bigg[  Y^2(s_i)\bigg]-E\bigg[  Y(s_i)Y(s_j)\bigg]+\frac{1}{2}E\bigg[  Y^2(s_j)\bigg]
\end{align}
Thus, 
\begin{equation}\label{eq1}
E\bigg[  Y(s_i)Y(s_j)\bigg]=-\gamma(s_i-s_j)+\frac{1}{2}\bigg(E\bigg[  Y^2(s_i)\bigg]+E\bigg[  Y^2(s_j)\bigg]\bigg)
\end{equation}
and similarly,
\begin{equation}\label{eq2}
E\bigg[  Y(s_0)Y(s_i)\bigg]=-\gamma(s_i-s_0)+\frac{1}{2}\bigg(E\bigg[  Y^2(s_i)\bigg]+E\bigg[  Y^2(s_0)\bigg]\bigg)
\end{equation}

Now, focusing on the mean squared-error, we have:

\begin{align}
E \bigg[ (Y(s_0)-\hat{Y}(s_0))^2\bigg] &= E \bigg[ \bigg(Y(s_0)-\sum_{i=1}^{85}\lambda_iY(s_i)\bigg)^2\bigg]= \sum_{i=1}^{85}\sum_{j=1}^{85}E\bigg[ Y(s_i)Y(s_j)\bigg]+\nonumber \\
& + E\bigg[ Y^2(s_0)\bigg]-2\sum_{i=1}^{85}\lambda_i E \bigg[ Y(s_0)Y(s_i)\bigg] \underbrace{=}_{\ref{eq1},\ref{eq2}} -\sum_{i=1}^{85}\sum_{j=1}^{85}\gamma(s_i-s_j)+\nonumber \\
&+\frac{1}{2}\sum_{i=1}^{85}\sum_{j=1}^{85}\lambda_i\lambda_j E \bigg[ Y(s_i)\bigg]+\frac{1}{2}\sum_{i=1}^{85}\sum_{j=1}^{85}\lambda_i\lambda_j E \bigg[ Y(s_j)\bigg]+E \bigg[ Y(s_0)\bigg]+\nonumber\\
&+2\sum_{i=1}^{85}\lambda_i \gamma(s_i-s_0)-\sum_{i=1}^{85}\lambda_i E\bigg[  Y^2(s_i)\bigg]-\sum_{i=1}^{85}\lambda_i E\bigg[  Y^2(s_0)\bigg]\underbrace{=}_{\sum_{i=1}^{85}\lambda_i =1}\nonumber \\
&=-\sum_{i=1}^{85}\sum_{j=1}^{85}\lambda_i \lambda_j \gamma(s_i-s_j)+ 2\sum_{i=1}^{85}\gamma(s_i-s_0) \nonumber
\end{align}

\newpage
\begin{landscape}
\section{Results of the temporal model}
\pagestyle{empty}%
\setlength{\tabcolsep}{2pt}
\begin{longtable}{c c c c c c c c c c c c c c c c c c c}

\caption{Estimated parameters of the temporal model in all grid points}\\
\hline
\textbf{Point} & \textbf{$a_0$} & \textbf{$a_1$} & \textbf{$a_2$} & \textbf{$a_3$} & \textbf{$a_4$} &\textbf{$a_5$} &\textbf{$a_6$}&\textbf{$a_7$}&\textbf{$a_8$}&\textbf{$a_9$}&\textbf{$a_{10}$}&\textbf{$a_{11}$}&\textbf{$a_{12}$} &\textbf{$\alpha_1$}&\textbf{$\alpha_2$}&\textbf{$b_0$}&\textbf{$b_1$}&\textbf{$b_2$}\\\hline
\hline
\endfirsthead
\multicolumn{19}{c}%
{\tablename\ \thetable\ -- \textit{Continued from previous page}} \\
\hline
\textbf{Point} & \textbf{$a_0$} & \textbf{$a_1$} & \textbf{$a_2$} & \textbf{$a_3$} & \textbf{$a_4$} &\textbf{$a_5$} &\textbf{$a_6$}&\textbf{$a_7$}&\textbf{$a_8$}&\textbf{$a_9$}&\textbf{$a_{10}$}&\textbf{$a_{11}$}&\textbf{$a_{12}$} &\textbf{$\alpha_1$}&\textbf{$\alpha_2$}&\textbf{$b_0$}&\textbf{$b_1$}&\textbf{$b_2$}\\\hline
\endhead
\hline \multicolumn{19}{r}{\textit{Continued on next page}} \\
\endfoot
\hline
\endlastfoot
1&1.3948&0.2057&0.1195&0.0022&-0.0532&0.0171&0.0320&-0.0227&-0.0146&-0.0215&0.0050&-0.0142&0.0121&0.5819&-0.0918&0.1324&0.0340&0.0099\\
2&1.4523&0.1744&0.0759&0.0405&-0.0390&-0.0081&0.0052&-0.0166&0.0110&-0.0203&-0.0013&-0.0130&0.0041&0.3668&-0.0501&0.0749&0.0493&0.0117\\
3&1.2287&0.2123&0.1321&-0.0197&-0.0443&-0.0051&0.0093&-0.0049&-0.0015&-0.0108&-0.0020&-0.0226&0.0020&0.5674&-0.1141&0.0990&0.0605&0.0126\\
4&1.7764&-0.0959&0.0785&-0.0359&-0.0118&0.0352&-0.0080&0.0100&-0.0161&0.0009&0.0090&-0.0126&-0.0132&0.6442&-0.1500&0.1370&0.0393&0.0070\\
5&1.4248&-0.0355&0.0723&0.0199&-0.0251&0.0150&-0.0007&0.0082&-0.0043&-0.0136&0.0057&-0.0090&0.0219&0.4543&-0.0844&0.0942&0.0368&0.0133\\
6&1.2190&0.1334&0.0794&-0.0320&-0.0283&-0.0303&0.0017&0.0257&-0.0025&-0.0213&0.0039&-0.0349&-0.0063&0.5074&-0.0790&0.0831&0.0418&0.0075\\
7&1.1497&0.0171&0.1457&-0.0732&-0.0336&0.0638&-0.0082&0.0086&0.0057&-0.0107&-0.0020&0.0050&0.0292&0.6112&-0.1029&0.1285&0.0531&0.0137\\
8&1.0077&-0.2526&0.0917&-0.0085&0.0315&0.0475&-0.0023&0.0176&-0.0104&-0.0052&0.0134&-0.0088&0.0023&0.4311&-0.1208&0.0829&0.0638&0.0164\\
9&1.0087&-0.1505&0.1246&-0.0261&0.0106&0.0278&0.0207&0.0231&0.0103&0.0017&0.0015&-0.0059&0.0148&0.5156&-0.0801&0.0963&0.0860&0.0274\\
10&0.9302&0.1483&0.1317&-0.0903&-0.0245&-0.0253&-0.0166&0.0626&0.0151&-0.0047&0.0086&-0.0287&-0.0082&0.5483&-0.1168&0.1662&0.0967&0.0233\\
11&1.0922&-0.2404&0.0784&-0.0300&0.0199&0.0318&0.0032&0.0336&0.0049&-0.0107&-0.0089&-0.0145&0.0038&0.3455&-0.0503&0.0843&0.0563&0.0169\\
12&0.7483&-0.1729&0.1057&0.0026&-0.0281&0.0425&-0.0024&0.0260&-0.0068&-0.0111&-0.0195&-0.0187&-0.0056&0.4458&-0.0724&0.1288&0.1057&0.0050\\
13&0.6495&0.1396&0.0432&-0.0160&-0.0113&-0.0137&-0.0087&-0.0031&0.0155&-0.0191&-0.0144&-0.0167&0.0026&0.4897&-0.0661&0.1076&0.1005&-0.0046\\
14&1.3966&0.1573&0.0658&0.0107&-0.0160&0.0264&-0.0219&-0.0083&0.0046&-0.0172&-0.0013&0.0061&-0.0005&0.3865&0.0000&0.1316&0.0892&0.0220\\
15&0.9573&-0.0980&0.0623&0.0394&-0.0225&0.0273&-0.0160&0.0079&-0.0015&-0.0152&-0.0059&0.0094&0.0021&0.4948&-0.1181&0.0949&0.0695&0.0019\\
16&1.5964&-0.0931&0.1040&0.0386&-0.0377&0.0218&-0.0284&0.0031&-0.0230&-0.0146&-0.0006&-0.0017&-0.0150&0.5000&-0.1466&0.1321&0.0780&-0.0016\\
17&1.4013&0.3464&0.0633&0.0414&-0.0856&-0.0103&-0.0259&0.0011&0.0246&-0.0455&-0.0324&-0.0316&-0.0194&0.5816&-0.1459&0.1491&0.1330&-0.0026\\
18&1.2473&0.0775&0.0874&-0.0314&-0.0272&0.0173&-0.0211&0.0071&0.0123&-0.0115&-0.0106&-0.0012&-0.0075&0.4781&0.0000&0.0706&0.0616&0.0160\\
19&1.4926&-0.3566&0.1319&-0.0621&-0.0253&0.0130&-0.0202&0.0215&-0.0282&0.0027&0.0046&0.0005&-0.0140&0.5781&-0.1898&0.1845&0.1547&0.0064\\
20&1.6302&-0.2842&0.1147&-0.0022&0.0291&0.0403&-0.0230&-0.0004&-0.0008&0.0010&-0.0241&0.0364&0.0047&0.5463&-0.1707&0.2035&0.1402&-0.0005\\
21&1.5082&-0.1110&0.0779&0.0197&-0.0244&0.0186&0.0009&-0.0185&0.0045&-0.0082&-0.0165&0.0047&-0.0091&0.4754&-0.1367&0.1487&0.1200&0.0002\\
22&1.1074&0.0132&0.0492&-0.0094&-0.0249&0.0158&-0.0049&0.0033&0.0203&-0.0120&-0.0031&0.0072&0.0092&0.4830&0.0329&0.0509&0.0436&0.0059\\
23&1.2691&0.0811&0.0818&-0.0432&-0.0014&0.0268&0.0042&0.0078&0.0013&-0.0322&-0.0126&0.0269&-0.0007&0.4079&-0.0526&0.1172&0.0664&0.0208\\
24&1.3424&-0.3635&0.1345&-0.0579&0.0446&0.0008&-0.0215&0.0021&-0.0271&0.0017&-0.0199&0.0329&-0.0030&0.5941&-0.2013&0.2586&0.2102&0.0038\\
25&1.7139&-0.3740&0.1331&-0.0370&0.0229&0.0255&-0.0181&0.0053&-0.0222&0.0117&-0.0143&0.0385&-0.0134&0.5793&-0.1874&0.2392&0.2175&-0.0120\\
26&1.2735&-0.0725&0.1357&-0.0528&-0.0337&0.0244&0.0268&0.0334&-0.0031&-0.0114&-0.0126&-0.0088&-0.0324&0.4959&-0.1299&0.1722&0.1097&0.0211\\
27&1.2066&-0.1219&0.0196&0.0031&-0.0301&0.0052&-0.0145&-0.0074&0.0178&-0.0154&0.0044&-0.0159&-0.0075&0.5085&0.0085&0.0674&0.0724&-0.0021\\
28&0.9431&0.0855&0.0800&-0.0166&-0.0310&0.0277&0.0285&-0.0149&0.0208&-0.0145&-0.0161&0.0102&0.0088&0.4969&-0.0672&0.0997&0.0960&0.0045\\
29&1.2316&0.1201&0.0743&-0.0092&-0.0128&0.0322&0.0167&-0.0263&0.0109&-0.0231&-0.0235&0.0238&0.0154&0.4560&-0.0536&0.1058&0.0588&0.0166\\
30&1.4165&-0.1705&0.1191&-0.0187&0.0208&0.0100&-0.0071&0.0057&-0.0106&-0.0174&-0.0008&0.0082&0.0036&0.5044&-0.1582&0.2131&0.1443&0.0103\\
31&1.6573&-0.2575&0.1136&-0.0281&-0.0031&0.0241&-0.0114&0.0075&-0.0288&-0.0078&-0.0104&0.0198&-0.0089&0.5801&-0.1872&0.1739&0.1430&0.0073\\
32&1.3092&-0.1071&0.1425&-0.0095&-0.0608&0.0293&-0.0039&-0.0090&-0.0398&-0.0297&-0.0007&-0.0007&-0.0029&0.5519&-0.1800&0.1563&0.0867&0.0156\\
33&0.8500&-0.0636&0.0453&-0.0065&-0.0331&0.0242&-0.0258&0.0115&0.0107&-0.0135&0.0088&0.0061&-0.0003&0.4232&0.0304&0.0867&0.0710&0.0093\\
34&1.0796&0.0892&0.0719&0.0062&-0.0435&0.0283&0.0063&-0.0077&0.0142&-0.0153&-0.0157&-0.0029&0.0156&0.4609&0.0000&0.0914&0.0925&0.0005\\
35&0.9888&0.1254&0.0683&0.0190&-0.0345&0.0338&0.0179&-0.0261&0.0164&-0.0272&-0.0263&-0.0034&0.0149&0.4620&0.0000&0.0911&0.0909&0.0088\\
36&1.1520&0.0933&0.0628&0.0119&-0.0192&0.0418&0.0126&-0.0247&0.0023&-0.0272&-0.0192&0.0185&0.0194&0.3952&0.0000&0.0816&0.0598&0.0104\\
37&1.4060&-0.1162&0.1312&-0.0140&0.0069&0.0106&0.0044&0.0125&0.0081&-0.0036&-0.0031&-0.0019&-0.0086&0.4982&-0.1575&0.2094&0.1397&0.0015\\
38&1.4985&-0.4075&0.1513&-0.0636&0.0098&0.0283&-0.0060&0.0153&-0.0296&0.0045&0.0004&0.0368&-0.0015&0.6282&-0.2155&0.1976&0.1590&0.0263\\
39&1.5247&-0.1895&0.1344&-0.0170&-0.0645&0.0317&-0.0179&0.0252&-0.0186&-0.0115&0.0043&-0.0045&-0.0266&0.4626&-0.1516&0.1598&0.0881&0.0283\\
40&1.2869&-0.2269&0.1040&-0.0766&-0.0549&0.0104&-0.0053&-0.0041&-0.0019&-0.0060&-0.0112&-0.0183&-0.0060&0.5770&-0.1700&0.1484&0.0997&0.0073\\
41&1.3865&-0.0555&0.1343&-0.0208&-0.0673&0.0258&0.0088&0.0091&-0.0144&-0.0073&0.0084&0.0218&-0.0010&0.6105&-0.1627&0.1496&0.0738&0.0188\\
42&1.4279&-0.0307&0.1386&-0.0380&-0.0616&0.0191&0.0044&0.0014&-0.0132&-0.0172&-0.0044&0.0240&0.0099&0.5669&-0.1602&0.1744&0.0917&0.0183\\
43&1.3167&0.3227&0.0638&0.0199&-0.0746&0.0082&-0.0048&-0.0241&0.0199&-0.0419&-0.0210&-0.0270&0.0167&0.5315&-0.1023&0.1561&0.1586&0.0023\\
44&1.6048&-0.1952&0.0558&-0.0046&-0.0016&0.0081&-0.0239&0.0075&0.0008&-0.0037&0.0082&-0.0070&-0.0088&0.4916&-0.1299&0.1302&0.0951&0.0036\\
45&1.0506&-0.2301&0.1432&-0.0776&0.0066&0.0052&0.0088&0.0082&0.0099&-0.0062&-0.0166&-0.0041&-0.0092&0.5175&-0.1428&0.1913&0.1440&0.0076\\
46&1.3975&-0.0106&0.0841&-0.0134&-0.0480&0.0080&0.0086&0.0118&0.0025&-0.0158&0.0000&-0.0273&-0.0297&0.4287&-0.1051&0.1447&0.1256&-0.0017\\
47&1.4279&0.0402&0.1000&-0.0194&-0.0184&0.0181&0.0033&0.0109&0.0363&-0.0074&-0.0018&-0.0168&-0.0038&0.4918&-0.0986&0.1635&0.1210&0.0031\\
48&1.1579&-0.2191&0.0595&-0.0926&-0.0193&-0.0322&0.0054&0.0128&0.0205&0.0060&-0.0208&-0.0265&-0.0180&0.4411&-0.0686&0.0929&0.0756&-0.0088\\
49&0.9093&-0.2278&0.1234&-0.1282&-0.0076&-0.0009&0.0095&0.0178&0.0113&0.0083&-0.0157&0.0024&0.0025&0.5094&-0.1553&0.1348&0.0836&0.0104\\
50&1.4485&-0.0867&0.1387&-0.0630&-0.0528&0.0275&0.0181&0.0128&-0.0205&-0.0022&-0.0176&0.0326&0.0116&0.5412&-0.1862&0.2024&0.1097&0.0356\\
51&1.1908&-0.1086&0.1040&-0.0503&-0.0185&-0.0058&0.0096&0.0019&-0.0070&-0.0114&0.0167&-0.0057&-0.0117&0.5127&-0.1474&0.1739&0.1066&0.0101\\
52&1.2444&-0.0535&0.1207&-0.0539&-0.0416&-0.0047&0.0131&0.0235&0.0186&-0.0124&0.0126&-0.0012&-0.0027&0.4913&-0.1349&0.1323&0.0857&0.0130\\
53&1.1151&-0.2214&0.0935&-0.0557&-0.0123&0.0038&0.0040&0.0189&0.0222&0.0108&-0.0220&-0.0080&-0.0074&0.4770&-0.1099&0.1677&0.0834&0.0035\\
54&1.3358&-0.0193&0.0737&-0.0050&-0.0322&0.0106&0.0052&0.0117&0.0126&-0.0170&-0.0111&-0.0159&-0.0170&0.4414&-0.1089&0.1001&0.0684&0.0052\\
55&1.4260&0.0516&0.0580&-0.0060&-0.0327&-0.0085&-0.0143&0.0093&0.0131&-0.0108&-0.0111&-0.0196&-0.0193&0.5255&-0.1494&0.1488&0.1187&-0.0025\\
56&1.2038&-0.2124&0.1215&-0.0725&-0.0302&0.0083&0.0232&0.0084&-0.0129&-0.0043&-0.0281&-0.0013&0.0021&0.5166&-0.1952&0.1562&0.0798&0.0228\\
57&0.9399&-0.0168&0.0451&-0.0384&-0.0165&-0.0094&-0.0207&0.0147&-0.0015&0.0011&-0.0059&-0.0208&-0.0059&0.3971&-0.1064&0.1489&0.0772&0.0252\\
58&1.7448&0.0388&0.0693&-0.0109&-0.0332&-0.0090&0.0121&0.0090&0.0117&-0.0311&0.0270&-0.0143&-0.0079&0.4513&-0.1220&0.1546&0.0784&-0.0014\\
59&1.6838&0.0684&0.0825&-0.0239&-0.0720&-0.0173&-0.0057&0.0173&0.0043&-0.0272&0.0124&-0.0332&-0.0148&0.4526&-0.1218&0.1570&0.1150&-0.0046\\
60&1.3975&-0.0304&0.0609&-0.0105&-0.0204&-0.0027&-0.0126&0.0096&0.0238&-0.0251&0.0042&-0.0273&-0.0171&0.4825&-0.1303&0.1803&0.1342&-0.0076\\
61&1.3881&0.0179&0.0623&0.0189&-0.0353&0.0036&-0.0113&0.0162&0.0142&-0.0202&-0.0148&-0.0195&-0.0320&0.4662&-0.1207&0.1528&0.0798&0.0123\\
62&1.6469&-0.0908&0.0734&-0.0186&-0.0410&0.0028&0.0031&0.0033&-0.0064&-0.0242&-0.0130&-0.0085&-0.0152&0.4710&-0.1332&0.1528&0.0955&0.0089\\
63&1.5859&0.0298&0.0880&0.0004&-0.0659&0.0075&-0.0077&-0.0120&-0.0081&-0.0105&-0.0080&-0.0281&-0.0082&0.4606&-0.1576&0.1603&0.0886&-0.0060\\
64&1.3990&-0.0308&0.0458&0.0059&-0.0283&-0.0059&-0.0119&-0.0087&0.0003&-0.0001&0.0014&-0.0217&-0.0049&0.3904&-0.1142&0.1103&0.0418&0.0076\\
65&1.5121&-0.0612&0.0476&0.0204&-0.0040&0.0044&0.0014&0.0067&-0.0016&-0.0340&-0.0025&-0.0105&-0.0141&0.4328&-0.1313&0.1469&0.0636&-0.0067\\
66&1.5670&0.0028&0.0395&0.0497&-0.0290&0.0027&-0.0093&-0.0051&0.0091&-0.0279&-0.0101&-0.0227&-0.0157&0.4911&-0.1513&0.1860&0.0947&-0.0053\\
67&1.3138&-0.0949&0.0770&0.0116&-0.0022&0.0110&0.0000&-0.0018&0.0207&-0.0362&-0.0101&-0.0264&-0.0217&0.5452&-0.1824&0.2074&0.1359&-0.0190\\
68&1.3033&-0.1380&0.0774&-0.0118&-0.0160&0.0093&-0.0073&0.0115&0.0022&-0.0387&-0.0125&-0.0165&-0.0342&0.5510&-0.1951&0.1620&0.1088&-0.0010\\
69&1.4821&-0.2025&0.0609&0.0044&0.0112&0.0034&-0.0024&-0.0052&0.0015&-0.0322&0.0014&-0.0042&-0.0205&0.4362&-0.1771&0.1374&0.0941&0.0011\\
70&1.5232&0.0312&0.0633&0.0398&-0.0410&0.0144&-0.0344&0.0006&0.0099&-0.0228&-0.0076&-0.0247&-0.0214&0.4916&-0.1747&0.1387&0.0769&-0.0016\\
71&1.5959&-0.0809&0.0114&0.0647&-0.0125&-0.0034&-0.0274&-0.0179&0.0074&-0.0206&0.0152&-0.0165&-0.0215&0.4643&-0.1327&0.1340&0.0478&0.0050\\
72&1.2104&-0.1449&0.0849&-0.0415&-0.0277&0.0127&0.0138&0.0103&-0.0006&-0.0360&0.0035&-0.0042&-0.0149&0.5725&-0.1874&0.1700&0.0702&0.0013\\
73&1.4962&-0.1020&0.0855&-0.0458&-0.0215&-0.0217&0.0304&0.0138&0.0152&-0.0277&0.0166&0.0001&-0.0113&0.5227&-0.1525&0.1494&0.0917&-0.0048\\
74&1.4492&-0.1065&0.0549&-0.0382&-0.0106&0.0064&0.0001&0.0149&0.0136&-0.0171&-0.0121&0.0009&-0.0269&0.4918&-0.1323&0.1086&0.0523&-0.0017\\
75&1.5120&-0.1180&0.0681&-0.0222&-0.0063&0.0062&-0.0012&0.0096&0.0115&-0.0223&-0.0069&0.0009&-0.0230&0.4844&-0.1613&0.1211&0.0591&-0.0027\\
76&1.5434&-0.1039&0.0815&-0.0174&-0.0079&0.0101&0.0064&0.0278&0.0025&-0.0181&0.0058&0.0062&-0.0166&0.4912&-0.1393&0.1152&0.0584&-0.0055\\
77&1.4776&-0.1061&0.1077&-0.0612&-0.0235&-0.0018&-0.0052&0.0140&0.0105&-0.0315&0.0083&-0.0151&-0.0013&0.5021&-0.1367&0.1338&0.1039&-0.0040\\
78&1.5154&-0.2336&0.0455&0.0239&0.0247&0.0152&-0.0186&0.0066&0.0042&-0.0245&0.0138&-0.0036&-0.0163&0.4711&-0.1320&0.1293&0.0642&0.0031\\
79&1.6220&-0.0101&0.0579&-0.0188&-0.0551&-0.0094&0.0129&0.0059&0.0019&-0.0158&0.0116&0.0121&0.0005&0.5373&-0.1529&0.1275&0.0476&0.0082\\
80&1.4980&0.0286&0.1003&-0.0272&-0.0616&-0.0090&0.0346&0.0063&-0.0077&-0.0314&0.0160&0.0062&-0.0150&0.4945&-0.1183&0.1152&0.0508&-0.0013\\
81&1.2417&-0.0813&0.0868&-0.0540&-0.0344&0.0013&0.0207&0.0052&0.0076&-0.0218&0.0083&-0.0003&-0.0162&0.4932&-0.1523&0.0922&0.0370&0.0054\\
82&1.2909&-0.0935&0.0764&-0.0409&-0.0306&0.0071&0.0127&0.0195&0.0051&-0.0134&0.0065&0.0086&-0.0152&0.5073&-0.1390&0.0961&0.0417&-0.0003\\
83&1.5420&-0.0715&0.0571&-0.0097&-0.0164&0.0142&-0.0032&0.0253&0.0021&-0.0153&0.0015&0.0043&-0.0129&0.4562&-0.1237&0.0948&0.0403&0.0033\\
84&1.5042&-0.0498&0.0708&0.0209&-0.0279&0.0053&-0.0037&0.0030&0.0065&-0.0304&0.0017&-0.0091&-0.0056&0.4619&-0.1464&0.1039&0.0663&-0.0091\\
85&1.4821&-0.2418&0.0479&0.0247&0.0089&0.0143&0.0018&-0.0042&0.0014&-0.0327&0.0148&-0.0025&-0.0160&0.5069&-0.1627&0.1346&0.0760&0.0148
\label{allparams}
\end{longtable}
\end{landscape}
\end{document}